\documentclass[%
 reprint,
nofootinbib,
 amsmath,amssymb,
 aps,
]{revtex4-1}

\usepackage{graphicx}
\usepackage{dcolumn}
\usepackage{bm}
\usepackage[colorlinks=true,citecolor=blue,urlcolor=blue]{hyperref}
\usepackage[mathlines]{lineno}

\usepackage{aas_macros}
\usepackage[dvipsnames]{xcolor}
\usepackage[normalem]{ulem}

\newcommand{\Mpch}{$h^{-1}\,\mbox{Mpc}$}
\newcommand{\kpch}{$h^{-1}\,\mbox{kpc}$\,}

\newcommand{\xiis}{$\xi(s_\perp,s_\parallel)$~}

\newcommand{\ie}{i.e.}
\newcommand{\eg}{e.g.}
\newcommand{\kps}{km\,s$^{-1}$\,}
\newcommand{\elephant}{\textsc{elephant}}
\newcommand{\lcdm}{$\mathrm{\Lambda CDM}$}
\newcommand{\de}{\mathrm{d}}
\newcommand{\etal}{\textit{et al.}}
\newcommand{\calr}{\mathcal{R}}

\def\simlt{\stackrel{<}{{}_\sim}}

\def\HI{$\mathcal{H}_1$}
\def\HII{$\mathcal{H}_2$}
\def\HIII{$\mathcal{H}_3$}
\def\GA{$\mathcal{G}$}

\begin{document}

\title{Probing gravity with redshift-space distortions:\\effects of tracer bias and sample selection}

\author{Jorge Enrique Garc\'ia-Farieta}
\email{(jorge,\,hellwing,\,gupta,\,bilicki)@cft.edu.pl}
\author{Wojciech A. Hellwing}%
\author{Suhani Gupta}%
\author{Maciej Bilicki}%
\affiliation{Center for Theoretical Physics, Polish Academy of Sciences, Al. Lotnik\'ow 32/46, 02-668 Warsaw, Poland }

\date{\today}

\begin{abstract}
We investigate clustering properties of dark matter halos and galaxies to search for optimal statistics and scales where possible departures from general relativity (GR) could be found. We use large N-body cosmological simulations to perform measurements based on the two-point correlation function (2PCF) in GR and in selected modified gravity (MG) structure formation scenarios. As a test-bed, we employ two popular beyond-GR models: $f(R)$ gravity and the normal branch of the Dvali-Gabadadze-Porrati (nDGP)  braneworld. We study a range of simulated halo and galaxy populations and reveal a noticeable MG signal in the monopole and quadrupole moments of the redshift-space 2PCF, and in the so-called clustering wedges. However, once expressed in terms of the linear distortion parameter, $\beta$, the statistical significance of these signals largely diminishes due to a strong degeneracy between MG-enhanced clustering and modified tracer bias. To circumvent this, we consider statistics less dependent on the bias: relative clustering ratios. We generalize the monopole ratio proposed in earlier work to multipole moments and clustering wedges, and introduce a new  estimator of the $\beta$ parameter. The clustering ratios we extract foster noticeable differences between MG and GR models, reaching a maximum deviation of 10\% at 2$\sigma$ significance for specific variants of $f(R)$ and nDGP. We show that such departures could be measured for $\beta$ if non-linear effects at intermediate scales are correctly modeled. Our study indicates that the clustering ratios give great promise to search for signatures of MG in the large-scale structure. We also find that the selection of an optimal tracer sample depends on a particular statistics and gravity model to be considered.
\end{abstract}

\maketitle


\section{Introduction}\label{sec:intro}

The standard cosmological model, Lambda-Cold Dark Matter (\lcdm), assumes general relativity (GR) as a description of gravity on all scales and at all times. A simple parametrisation of \lcdm{} in terms of just 6 quantities provides an excellent fit to numerous
observational data of various kinds collected over decades \cite[\eg][]{Semboloni2006,Hamuy2006PASP,Miknaitis2007ApJ,Conley2011ApJS,sdss_boss_2017,Planck_Legacy_2018}, but we are left with a puzzling outcome. According to \lcdm, the Universe, for the most of its evolution, is dominated by dark components of yet unknown physical nature:
collisionless cold dark matter (DM), constituting most of the cosmic mass, and cosmological constant as the source of dark energy (DE), that causes the observed acceleration of the cosmic expansion. The existence of the cosmological constant, commonly associated to a negative-pressure fluid, would have profound implications for fundamental physics, however, there is no compelling direct evidence for it yet \cite{Peebles2002DE,Sarkar2007DE,Frieman2008,Brax2017DE_review}.

The urgent need to explain the physical mechanism behind the cosmic acceleration and, more generally, the nature of the dark sector, gives motivation to investigate the validity of GR on cosmological scales and to consider models of gravity beyond Einstein's theory. Such alternative frameworks may postulate for instance modifications in the theory of gravity on cosmological scales, but to be viable they must be consistent with experiments. In particular, they need to predict the same, observationally well-constrained expansion history as \lcdm, and at the same time have to pass the stringent local and high-energy tests of gravity where GR has been shown to work very well \cite{Kramer2006,Uzan2011, Will2014,Collett2018,Abbott_TestsGRLIGO}. Among the proposed modifications of GR that we will refer to as `modified gravity' (MG), sometimes referred in the literature as Extended Theories of Gravity \cite[for a detailed review see \eg][]{Capozziello2011_ExtendedGravityTheories},
the models based on $f(R)$ gravity and braneworld stand out because of their generality and rich phenomenology \cite{koyama2007b,Sotiriou2010,DeFelice2010,Maartens2010}.

The large-scale structure of the Universe is one of the richest sources of cosmological information, available chiefly in the clustering of matter tracers. Clustering is an important probe of the underlying cosmological model in which cosmic structures evolve, and therefore, it can be used to test the gravity theory \cite{hamilton_review_1998,Scoccimarro_review_2004,Taruya2014TestGravityPhRvD,Ishak2019LRR,Ferreira2019ARA,BakerRevModPhys2021}. In fact, the growth rate of structures, a parameter that describes the time evolution of matter overdensities, can be used to differentiate from \lcdm\ those alternative MG models that otherwise have expansion history compatible to GR.

One of main statistical measures of clustering is the two-point correlation function (2PCF), or the power spectrum, its equivalent in the Fourier space. In this work we study the former of these
statistics, with the main focus on the redshift space, in which a wealth of observations are made. We also investigate the tracer (galaxy and halo) bias, which is notoriously difficult to model and
constitutes one of the main sources of systematic errors plaguing the derivations of cosmological parameters from galaxy redshift surveys \cite[\eg][]{Desjacques2018, Verde_bias2002MNRAS}.

A specific feature suitable to test gravity models using the galaxy distribution, and in particular to measure the growth rate, is redshift-space distortions (RSD; \cite{Jackson_FoG_1972,kaiser_1987}). These appear naturally in observations due to radially-projected peculiar velocities that are a direct consequence of the growth of structure and the gravitational interaction. RSD induce anisotropy in the observationally derived 2PCF and are directly accessible in surveys that provide sky positions and spectroscopic redshifts of the tracers.

In view of the forthcoming and planned redshift surveys offering much better statistics than possible so far \cite{Laureijs_2011,DESI}, challenges arise in using RSD for robust cosmological constraints and reliable tests of gravity. In particular, modeling of RSD is a paramount task, but not always successful, in the estimation of the growth factor. Several RSD models commonly extended from the linear theory formalism have proven to perform well at large scales, however, exploration of the nonlinear regime, where the most cosmological information is contained, is crucial in testing gravity. The commonly used modeling generally fails to describe accurately the full dynamics of the tracers at non-linear and weakly non-linear scales \cite{peebles1980large,Fisher1995RSD_Streaming}.
Over time, more elaborated approaches than the standard linear dispersion model have been proposed based on different perturbation theory schemes \cite{Scoccimarro_review_2004,TNS_model,Bella2017JCAPRSD,Song2018JCAPRSD}. In general, however, these more advanced approaches were not designed for nor thoroughly tested on alternative gravity scenarios.
Only in the recent years some progress in that aspect has been made, but the developments apply only to RSD of a smooth DM component, rather than to the full clustering
of biased tracers, and furthermore they are available only for some selected MG models \cite{KoyamaPhysRevD79,Taruya2014TestGravityPhRvD,Bose2016,Taruya2016,Bose2017}.

The difficulties described above are often combated thanks to high-resolution N-body simulations, which by construction can probe
deeply into the non-linear regime. They thus provide both a means to test the performance of various RSD models, and a way to better understand
the galaxy biasing and velocities in the full non-linear regime. Running and employing N-body simulations for RSD studies has therefore become routine
practice in the field. However, high-resolution and fully non-linear simulations of MG are much more computationally expensive -- by a factor a few to even one order of magnitude -- than the same resolution \lcdm\ setup. As a consequence, MG simulations of volumes and resolutions sufficient for galaxy and halo RSD studies are usually prohibitively expensive. However, with the recent advent of improved and accelerated algorithms to solve for the additional MG physics, suitable N-body simulations became available for beyond-GR models \citep[\eg][]{Barreira2015JCAP,Bose2017JCAP,Arnold2019B,Winther2017,Arnold2019A}. Thanks to this, systematic studies comparing RSD signals and models in \lcdm\ vs.\ MG scenarios have become possible
\cite[\eg][]{Jennings_RSD_MG_2012,Marulli_2012B,Wyman2013,Barreira2016,DESI_ELEPHANT}.

In this paper, we investigate DM halo and galaxy clustering and the underlying growth rate of cosmic structures in two beyond-GR scenarios: $f(R)$ and nDGP, and compare them with \lcdm. For that purpose we employ a set of mock catalogs with number densities characteristic to those accessible in today's and near-future spectroscopic redshift surveys, such as SDSS, BOSS or DESI. These catalogs are based on a suite of state-of-the-art N-body simulations: the Extended LEnsing
PHysics using ANalaytic ray Tracing project (\elephant; \cite{Cautun2018MNRAS,DESI_ELEPHANT}). Our study is a continuation
and extension of earlier works that employed these simulations.
In Ref.~\cite{Arnalte-Mur2017MNRAS} DM halo populations of \lcdm\ and $f(R)$ models were studied
with a novel clustering statistic called the halo clustering ratio, $\mathcal{R}$.
That analysis indicated that the clustering ratio has potential of discriminating between standard gravity and the $f(R)$ models, especially those with a milder departure from GR.
More recently, Ref.~\cite{Aguayo2019oxg} investigated the clustering of \elephant\ mock galaxies using multipole moments and clustering wedges of the 2PCF. Their results confirm that the RSD measurements can help to distinguish between different gravity models. However, they highlight that the distortion parameter obtained $\beta$ with the linear RSD model is significantly underestimated for all the MG models they studied.

In our work we go beyond
the scope of these previous studies in a number of important aspects. First,
from the N-body simulations
we use mock galaxies as well as three different halo populations, which allows us to study the importance of tracer sampling density
in searching for MG signatures and analyze the behavior of the linear bias in various gravity models.
Second, we generalize the definition of the relative clustering ratio proposed in Ref. \cite{Arnalte-Mur2017MNRAS} to the multipole moments and clustering wedges, and introduce a new estimator based on
this ratio to constrain the linear distortion parameter, $\beta$, from redshift-space anisotropies. Although we only use the linear theory to model RSD, the new estimator is powerful enough to expose the impact of the scale at which measurements are made on recovering $\beta$. We also shed light on the degeneracy between the tracer bias and the growth rate, both encoded in the $\beta$ parameter. Finally we alleviate the issue addressed in \cite{Aguayo2019oxg} related to the efficacy of the linear model to recover the linear distortion parameter. We show that an appropriate estimator of $\beta$ based on linear theory can successfully recover its true value in the linear regime.

The paper is organized as follows: in Section
\ref{sec:modeltheory} we briefly describe the cosmological models studied here. In Section \ref{sec:simulations} we present the set of N-body simulations that have been employed in the analysis,
and provide details on the selected DM halo populations,
as well as the simulated galaxy sample. In Section \ref{sec:clustering} we discuss the estimators used to characterize the clustering measurements of different samples in both the real and redshift space. In Section \ref{sec:results} we present the results and analysis of the clustering measurements for different halo populations and the galaxy sample;
we also study the linear halo and galaxy bias in that Section.
Then, in Section \ref{sec:RelClustering} we introduce the estimator for relative clustering of multipole moments and wedges of the 2PCF, and discuss the results obtained for our samples. Finally, in Section \ref{sec:discussion_conclusions} we conclude and discuss the implications of our findings.


\section{Modified gravity models}
\label{sec:modeltheory}

Beyond-GR alternatives to the \lcdm\ cosmological model are numerous
(see \eg~the reviews \cite{Clifton2012,Joyce2015}), but numerous are also the problems they have to struggle with. Some
of these MG models are plagued with theoretical instabilities, and all of them obviously
have to face observational constraints, which often requires
fine-tuning of model parameters. One particularly simple extension of GR is by including a single scalar field, $\varphi$, in
the Einstein-Hilbert Lagrangian which sources gravity. However, when coupled to matter, the scalar field gives rise to
an additional gravitational force -- often referred to as the \textit{fifth force}. Such a fifth force can be quantified by
$\gamma\equiv |F_5|/F_N$,
where $F_N$ is the ‘standard’ Newtonian gravitational force that we obtain in the weak-field limit of GR.
Several experiments \citep[\eg][]{Creminelli2017,Baker2017} have constrained $\gamma\ll1$ on Earth, in the Solar System and in neutron star
binaries and merging black holes. This leaves two possibilities: either there is no room to deviate from GR on all scales, \ie{}
$\gamma \equiv 0$, or $\gamma$ is not a constant but instead varies in space (and possibly time). The latter scenario can be relatively
easily realized if the fifth-force arises due to the propagation
of extra degrees of freedom of a dynamical field that varies in space and time.
Models where this is the case are dubbed
\textit{screened} MG models, since the fifth force is screened in high-curvature or high-potential regions.

In this work we study two classes of MG models: the $f(R)$ gravity of Hu-Sawicki formulation \cite{Hu_Sawicki_2007} and the nDGP braneworld gravity.
In the former, the MG phenomenology arises from a generalization of the Ricci's scalar, $R$,
in the Einstein-Hilbert action to a functional of it, $f(R)$, and the fifth-force is manifested due to an extra degree of freedom
of the scalaron field of $R$. In the latter, the so-called normal branch of the Dvali-Gabadadze-Porrati (nDGP) model \cite{Dvali2000PhLB}, the fifth-force
is a manifestation of an extra degree of freedom that is due to bending of a 4-dimensional brane in the 5-dimensional
bulk space-time. In both cases, the arising fifth-force can be modeled as a gradient of an underlying phenomenological
scalar field.

We choose the $f(R)$ and nDGP gravity models since they constitute a very good test suite for a wider class of MG theories.
This is because most of the viable MG models can be divided into two general categories, depending on the physical mechanism of the fifth-force screening
they invoke. The screening can be either \textit{environment dependent} or \textit{object-mass dependent}.
The former responds to the local value of
the gravitational potential, and in the latter (also called the Vainshtein mechanism \cite{Vainshtein1972,Babichev2013usa}), the effectiveness
of the screening is usually moderated by
the local curvature of a given region of space. What is also essential here is that the two classes of MG models we consider pass the stringent
tests from the first Kilonova gravitational wave event of GW170817 \cite{Ezquiaga2017ekz}. Below we give a short description of the specific formulations
and settings of these two models. From the point of view of the statistics and observables studied in this paper, the most important
characteristic of our models is that we choose specific formulations of $f(R)$ and nDGP gravity theories that follow closely the \lcdm{}
expansion history. Thus the first non-zero physical effect that emerges from their extra degrees of freedom is imprinted
in modified history of the growth of structures.

\subsection{Dvali-Gabadadze-Porrati model}
The Dvali, Gabadadze \& Porrati (DGP) model \cite{Dvali2000PhLB} is inspired by string theory and assumes the existence of a 4+1-dimensional Minkowski space, within
which the ordinary 3+1-dimensional Minkowski space is embedded. In other words, the DGP model is one of the possible braneworld cosmology models,
where the Universe is described by a 4D brane which is embedded in a higher-dimensional spacetime called the bulk \cite{Sahni2003JCAP}. In
this respect, the so called normal branch DGP (nDGP) gravity is a natural extension of the DGP model \cite{Dvali2000PhLB, koyama2007ghosts} that provides an
explanation why the force of gravity is much weaker compared to the other fundamental forces \cite{Maartens2010}.
This is possible because fundamental matter particles are assumed to be confined to the brane, while gravity can propagate
through the extra spatial dimension(s).

The nDGP model introduces a free-parameter, the so-called crossing-over scale, $r_c$. It characterizes the scale at which
the 4-dimensional gravity of the brane ``leaks out'' to the 5-dimensional bulk space-time. This scale is simply obtained
as half of the ratio of the 5-dimensional Newton constant $G^{(5)}$ to the usual 4-dimensional one, denoted here simply as $G$:
\begin{equation}
r_c = \frac{1}{2} \frac{G^{(5)}}{G}.
\end{equation}
The crossing-over scale limits the size of the fundamental perturbation of the embedded 4D brane (\ie{} the maximum bending mode),
and the extra degree of freedom of the brane can be expressed by a free scalar field, $\varphi$.
On the linear level this translates to a maximum enhancement of the growth rate of structures, compared to the usual \lcdm{} background.
In the limit where the time derivatives of this new scalar field are negligible compared to its spatial derivatives, that is in
the so-called quasi-static limit \cite{koyama2007ghosts}, the model admits a modified Poisson equation for gravity
\begin{equation}
 \label{eqn:nDGP-Poisson}
 \nabla^2\Psi = 4\pi G a^2\rho\delta + \frac{1}{2} \nabla^2\varphi\,,
\end{equation}
where $\Psi$ is the classical Newtonian potential, $G$ is the 4D Newton constant, $a$ is the scale factor, $\rho$ is the matter background
density with its local density contrast $\delta$, and finally $\varphi$ is the scalar field
describing the extra degrees of freedom of the model. The new scalar field obeys its own equation of motion \cite{Schmidt2009PhRvD, Barreira2015JCAP, Winther2015PhRvD}
\begin{multline}
 \label{eqn:nDGP_scalar_eom}
 \nabla^2 \varphi + \frac{r_c^2}{3 \mathcal{B}(a) a^2}\left[\left(\nabla^2 \varphi\right)^2 -\left(\nabla_i \nabla_j \varphi\right)\left(\nabla^i \nabla^j \varphi\right)\right] = \\
	= \frac{8 \pi G a^2}{3 \mathcal{B}(a)} \rho \delta\,.
\end{multline}
The new function $\mathcal{B}(a)$ is defined as
\begin{equation}
 \label{eqn:nDGP_beta_function}
 \mathcal{B}(a) = 1 + 2 H r_c \left(1+ \frac{\dot{H}}{3 H^2} \right)\,.
\end{equation}
Here $H\equiv H(a)$ is the usual Hubble function of the background model. It is convenient now to define a dimensionless parameter, $\Omega_{rc}$,
which once specified will determine any given nDGP model. Thus, we define
\begin{equation}
 \label{eqn:omega_rc}
 \Omega_{rc}\equiv\frac{1}{\left(2r_c H_0\right)^2}\,,
\end{equation}
with $H_0$ denoting the present-day value of the Hubble parameter, and the above quantities are expressed assuming $c=1$.
In the phenomenological formulation presented above, the nDGP model admits at large scales a constant enhancement to the Newtonian
gravity, which can be evaluated in terms of an effective Newton constant,
\begin{equation}
G_{eff} = G \{1+ 1/[3 \mathcal{B}(a)]\}\,.
\label{eqn:G_eff}
\end{equation}
This therefore leads to a constant enhancement in the
linear-theory growth rate of structure, $f$,
by a factor
$\Delta f =G_{eff}/G$. On the smaller, non-linear scales the enhancement of gravity is effectively suppressed by the means
of the Vainshtein screening \cite{Vainshtein1972,Li2013}. We will study two variants of the nDGP model, specified by $r_c H_0=5$ ($\Omega_{rc}=0.01$),
and $r_c H_0=1$ ($\Omega_{rc}=0.25$), which we dub N5 and N1 respectively.

\subsection{$f(R)$ gravity}

The $f(R)$ gravity model is an extension of GR that has been extensively studied in the literature in the past several years
(see \eg{} \citep{Sotiriou2006} for a detailed review). Some previous works have explored this model in different and alternative contexts, such as the internal properties of cosmic structures and their mass-to-light ratio \cite[see \eg][]{Capozziello2009_ModelClustersfR,Salzano2014_scalarfield}, signatures of $f(R)$ gravity from thermodynamic equilibrium of the clustering of galaxies \cite{Capozziello2018_GalClusteringfR} as well as phenomenological scenarios than the chameleon, symmetron, and $f(R)$ gravity models that scale the local properties of astrophysical systems \cite{Salzano2017_Vainshtein}. The theory is obtained by substituting the Ricci scalar, $R$, in the Einstein-Hilbert action with an algebraic function $f(R)$. Here the accelerated expansion of the Universe is produced by this extra term replacing $\Lambda$ in the action integral,
without the need for any form of dark energy. The resulting modified theory of gravity is characterized by highly non-linear equations of
motion for the scalar field, and environment-dependent fifth-force screening is obtained via the so-called {\it chameleon mechanism} \cite{Sotiriou2010,Khoury2003PRD}.
The presence and effectiveness of the chameleon effect is very important for the viability of the $f(R)$-class theories. We comment more on this later on.

One particularly interesting and useful formulation
is the so-called Hu \& Sawicki \cite{Hu_Sawicki_2007} branch. Here the functional form of $f(R)$ is
\begin{equation}
 \label{eqn:fofR_func}
 f(R) = -m^2\frac{c_1\left(-R/m^2\right)^n}{ c_2\left(-R/m^2\right)^n+1}\,,
\end{equation}
where $n>0$, $c_1$ and $c_2$ are dimensionless free model parameters, and $m$ is an extra mass-dimension parameter.
Now, the extra degree of freedom can be again expressed in terms of a scalaron field, $f_R\equiv \de f(R)/\de R$, which is not massless,
unlike in the case of nDGP. We can relate the model parameters by writing \cite{Hu_Sawicki_2007}
\begin{equation}
 \label{eqn:f_r_combined}
 f_R=-n \frac{c_1}{c_2^2}\frac{\left(-R/m^2\right)^{n-1}}{ \left[1+\left(-R/m^2\right)^n\right]^2}\,,
\end{equation}
where the mass scale $m$ is defined as $m^2 \equiv H_0^2 \Omega_{\rm m}$.
For this $f(R)$ model, the background expansion history becomes consistent with the \lcdm\ case by choosing $c_{1}/c_{2} = 6\Omega _{\Lambda}/\Omega _{\rm M}$.
The scalaron field $f_{R}$ adds an additional degree of freedom to the model,
whose dynamics in the limit of $|f_R| \ll 1$ and $|f/R|\ll 1$ can be expressed in terms of perturbations of the scalar curvature,
$\delta R$, and matter overdensity, $\delta\rho$:
\begin{equation}\label{eqn:modPoisson}
\nabla^2 f_R = \frac{1}{3}\left(\delta R - 8 \pi G \delta \rho \right)\,.
\end{equation}

Comparing to the \lcdm\ model expansion history, and under the condition $\smash{c_2 \left(R/m^2\right)^n \gg 1}$,
the scalaron field can be approximated by:
\begin{equation}\label{eqn:fR-R_relation}
f_R \approx -n \frac{c_1}{c_2^2}\left(\frac{m^2}{-R}\right)^{n+1}\,.
\end{equation}
Consistency experiments on local gravity such as Solar-System constraints, as well as weak and strong equivalence principles constraints,
have been set the bound $n>0.5$ \cite[see \eg][]{Khoury2003PRD,Capozziello2008_SolarfRconstraintschameleon}.
In this paper we consider $f(R)$ models with $n=1$, which are consistent with the aforementioned constraints.

By setting $n=1$, the model is fully specified by only one free parameter, $c_2$, which in turn can be expressed in terms of
the dimensionless scalaron at present epoch, $f_{R0}$, given by:
\begin{equation}
\label{fR0}
f_{R0}\equiv -\frac{1}{c_{2}}\frac{6\Omega _{\Lambda }}{\Omega_{\rm m}}\left( \frac{m^2}{R_{0}}\right) ^{2}\,.
\end{equation}
Therefore, a particular choice of $f_{R0}$ fully specifies the
Hu-Sawicki $f(R)$ model. In this work we focus on the cases of
$f_{R0}=\{-10^{-5},-10^{-6}\}$, referred to from now on as F5
and F6, respectively.

At the background level the $f(R)$ theory can produce a significant fifth-forces at small non-cosmological scales. Presence of such forces
is however tightly constrained by Solar System and strong-field regime observational tests
\citep{Capozziello2008_SolarfRconstraintschameleon,Burrage2016,Sakstein2017}. Here, an essential
role is played by the already mentioned
chameleon mechanism, which effectively suppresses the fifth-force in high density regions. This intrinsically non-linear mechanism traps the scalar field
in high curvature regions making it very massive and suppressing deviations from GR dynamics. There are potentially many interesting non-linear effects
related to the chameleon mechanism that can affect stellar and galaxy evolution\citep[see \eg][]{Burrage2018}.
Here, we focus on cosmological scales and galaxy clustering, and so the simulations we employ implement self-consistently
the chameleon mechanisms only for dark matter clustering in the cosmological context.

\section{Halo and galaxy mock catalogs}\label{sec:simulations}

In this work we consider simulated cosmological data originating from
\elephant,
 introduced in
\cite{Cautun2018MNRAS,DESI_ELEPHANT}.
This is based on a suite of dark-matter-only
$N$-body simulations of the standard cosmological model, \lcdm, and of the two families of modified
gravity theories described above, nDGP and $f(R)$. The \elephant\ simulations have been run using the \textsc{ecosmog}
code \cite{Li2012JCAP}.

The simulations followed the dynamical evolution of $1024^3$
particles placed in
a box of $1024$\Mpch\ comoving width.
The evolution of DM phase-space was traced from the initial redshift $z_\mathrm{in}=49$ down to $z=0$,
with the comoving mass resolution of $m_p=7.798\times10^{10}M_{\odot}h^{-1}$ and comoving
force resolution of $\varepsilon=15$\kpch
equivalent of Plummer softening.
For each model we consider five independent phase realizations of the initial power spectrum,
which is derived for the \lcdm{} model with the WMAP9 collaboration best-fit parameters \cite{hinshaw2013nine}:
$\Omega_m=0.281$ (total fractional non-relativistic matter density), $\Omega_b = 0.046$ (fractional baryonic matter density),
$\Omega_\nu = 0.0$ (fractional relativistic matter species density), $\Omega_{\Lambda}=1-\Omega_m$ (cosmological constant
energy density), $\Omega_k=0$ (fractional curvature energy density), $h=0.697$
(Hubble constant in units of 100 \kps\,Mpc$^{-1}$),
$n_s = 0.971$ (primordial power spectrum slope), and $\sigma_8 = 0.820$ (linearly extrapolated
\lcdm{} power spectrum normalization).

In this work we will deal with DM halos and mock galaxies derived from the \elephant\ simulations,
saved at three cosmic epochs,
corresponding to redshifts $z=0,~ 0.3$ and $0.5$.
This particular redshift range is very interesting from both observations and theory vantage points.
The currently available redshift survey data, such as the BOSS LOWZ \cite{Cuesta2015mqa} attain the highest galaxy number density in this regime. These redshift ranges will be also probed by future, richer data from planned or already ongoing surveys like DESI \cite{DESI}, or 4MOST Cosmology Redshift Survey \cite{4HS_Richard2019dwt}.
Also, from the theoretical point of view, both the $f(R)$ and nDGP models exhibit the biggest
deviations
from \lcdm{} clustering at low and intermediate redshifts, $0.3\leq z\leq 0.7$ \citep[\eg][]{Hellwing2013, Cataneo2016JCAP,Barreira2016,Hellwing2017, Arnold2019B,Devi2019swk_HaloGal,Hellwing2020,Liu2021weo_fRDGP}.

The DM halos were identified in the simulations using the
\texttt{ROCKSTAR}\footnote{\url{https://bitbucket.org/gfcstanford/rockstar}}
\cite{Behroozi2013ApJ} halo finder. It is important to note that on top of using the friends-of-friends
approach in 6D phase-space, the halo finder performed additionally the gravitational unbinding procedure in which unbound particles are removed from a halo
in iterative steps.
The standard implementation of this procedure assumes Newtonian gravity for computing the potential and
particle binding energy. In $f(R)$ and nDGP, however, we can arrive at a situation where some halos
will be fully or partially unscreened, thus the total fifth-force should be added to their binding energy budget.
\texttt{ROCKSTAR} does not consider this extra binding energy in its calculations.
However, ignoring this extra contribution presents a conservative approach to unbinding:
if a particle were not bound in modified gravity, then it would be also definitely
unbound in \lcdm. In addition, the fraction of particles for which neglecting the extra
fifth-force potential would be significant is very small.

It has been shown that for all practical purposes relevant for this study one can neglect MG effects in the case of
$f(R)$ theories and use the standard Newtonian unbinding procedure \cite{Cautun2018MNRAS,Li2010PhRvD,Aviles2020JCAP}. For the nDGP case, the Vainshtein radius is larger than the biggest objects in our simulations for all the redshifts concerned \cite{koyama2007b,schmidt2009cluster}, which means that most of the halos should be self-screened.

The galaxy mock catalogs were built using a halo occupation distribution (HOD) prescription \cite{Berlind2002rn, Zheng2004id, Zheng2007zg}.
The HOD model parameters were tuned for each of the gravity models
independently to obtain a catalog matching the target number density and the projected real-space
clustering of BOSS-CMASS galaxies \cite{Manera2012sc}. These galaxy mock catalogs have been exploited in different works, such as \cite{Cautun2018MNRAS, Aguayo2018MNRAS,Aguayo2019oxg, Paillas2019,DESI_ELEPHANT}. In our analysis, we only consider central galaxies which are located at the center of potential of their host halo. For more details please refer to the original catalog release paper \cite{DESI_ELEPHANT}.

In our analysis we adapt $M_{200c}$ as our main halo mass definition.
This is defined as the DM mass enclosed within a sphere around the halo center,
with the radius $r_{200c}$, at which the spherically averaged density inside drops
down to a value $200$ times the critical density of the Universe, i.e. $\rho_c\equiv3H^2/8\pi G$.
Thus, whenever we refer to a halo mass, we mean $M_{200c}$, unless clearly stated otherwise.
Following Ref. \cite{Arnalte-Mur2017MNRAS}, from all the raw halo catalogs
we keep for further analysis the halos with at least 64 DM particles.
This sets our minimal halo mass to $M_{min}=5\times10^{12}\,M_\odot h^{-1}$. An important notice here is the fact that \elephant\ set is a pure N-body run, and as such,
treats all baryonic component as additional collisionless mass. The inclusion of highly non-linear baryonic physics
is fundamental for a proper modeling and understanding in full the galaxy formation process.
However, the impact of the full baryonic hydrodynamics for the halo and galaxy clustering and their peculiar velocities
has been shown to be minimal \cite[see \eg][]{Hellwing2016}. Thus, for our purposes here exclusion of any baryonic physics modeling should not affect our results.

\subsection{Halo mass function and selection of halo populations}
\label{sec:massfun}

The number density of tracers, which in the real survey situation are galaxies, for a volume-limited sample
can be simply related to the underlying halo mass function (HMF). The HMF quantifies the comoving number density of DM halos as a function of their mass for a given redshift and cosmology. Since we observe galaxies rather than their host halos, a usual approach is
to apply a chosen method of galaxy modeling to obtain the final mock galaxy catalog. In this work in addition to mock galaxies,
we also consider halo populations as tracers. This exercise will allow us to study the modified gravity signal
as a function of varying tracer number density.
In what follows we use the halo mass function to select our halo samples choosing their number
density to reflect some realistic observational values when using data from the current and next generation of galaxy surveys.

In standard cosmology HMF can be modeled by, to a high-degree universal, halo multiplicity function giving the number of collapsed objects as a function of the mass field variance
expressed at a corresponding scale. In modified gravity the fifth-force increases the mass field variance compared
to \lcdm, and so we can expect that HMF in MG will exhibit differences compared to the fiducial case
\cite{Schmidt2010PhRvD,Lam2012MNRAS,Clifton2012,Lombriser2013wta,Joyce2016,Hagstotz2018onp,garciafarieta2020massive}. In the cumulative form, the number density of halos above a threshold is given by
\begin{equation}
n\left(>M_{200 c}\right)=\int_{M_{min}}^{\infty} \frac{\mathrm{d}
n}{\mathrm{d} M_{200 c}} \mathrm{d} M_{200 c}.
\end{equation}
Here, as a threshold we choose a halo mass. The object under the above integral is the differential mass function (dMF), $d n (M,z)/ d M$, which encodes the number of halos per mass interval.
\begin{figure}
 \centering
 \includegraphics[width=\linewidth]{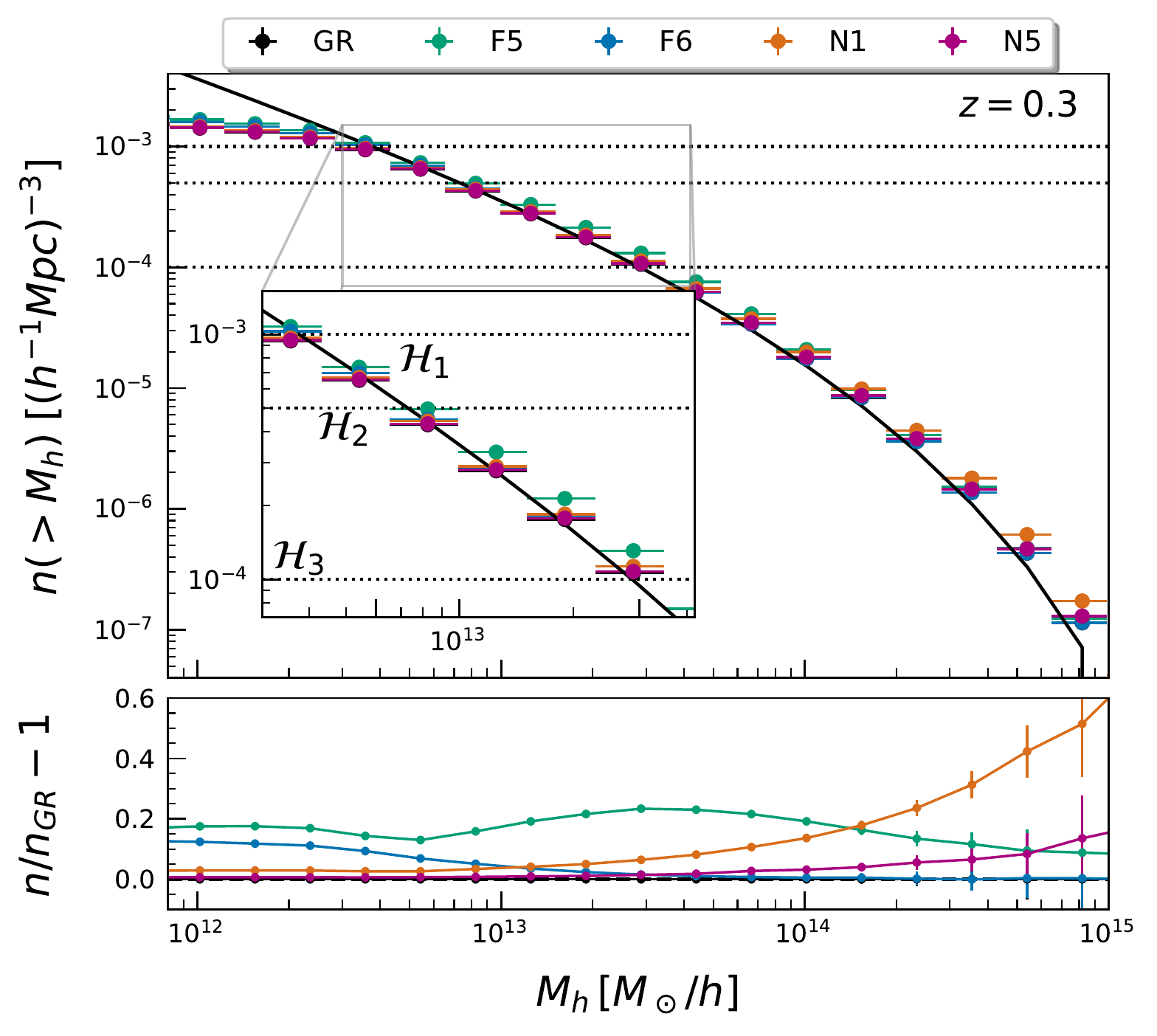}
 \caption{
 Cumulative halo mass function for the different \elephant\
 MG models taken at redshift $z = 0.3$. The
horizontal dotted lines signal the number densities we use to define
our three halo samples. The black solid line represents the theoretical expectation, for \lcdm, given by Tinker \etal\ \cite{Tinker2008}. The bottom panel shows the relative
deviation with respect to the \lcdm\ (GR) model with lines connecting the datapoints to guide the eye.}
 \label{fig:MF_papercum}
\end{figure}

\begin{figure*}
 \includegraphics[width=\linewidth]{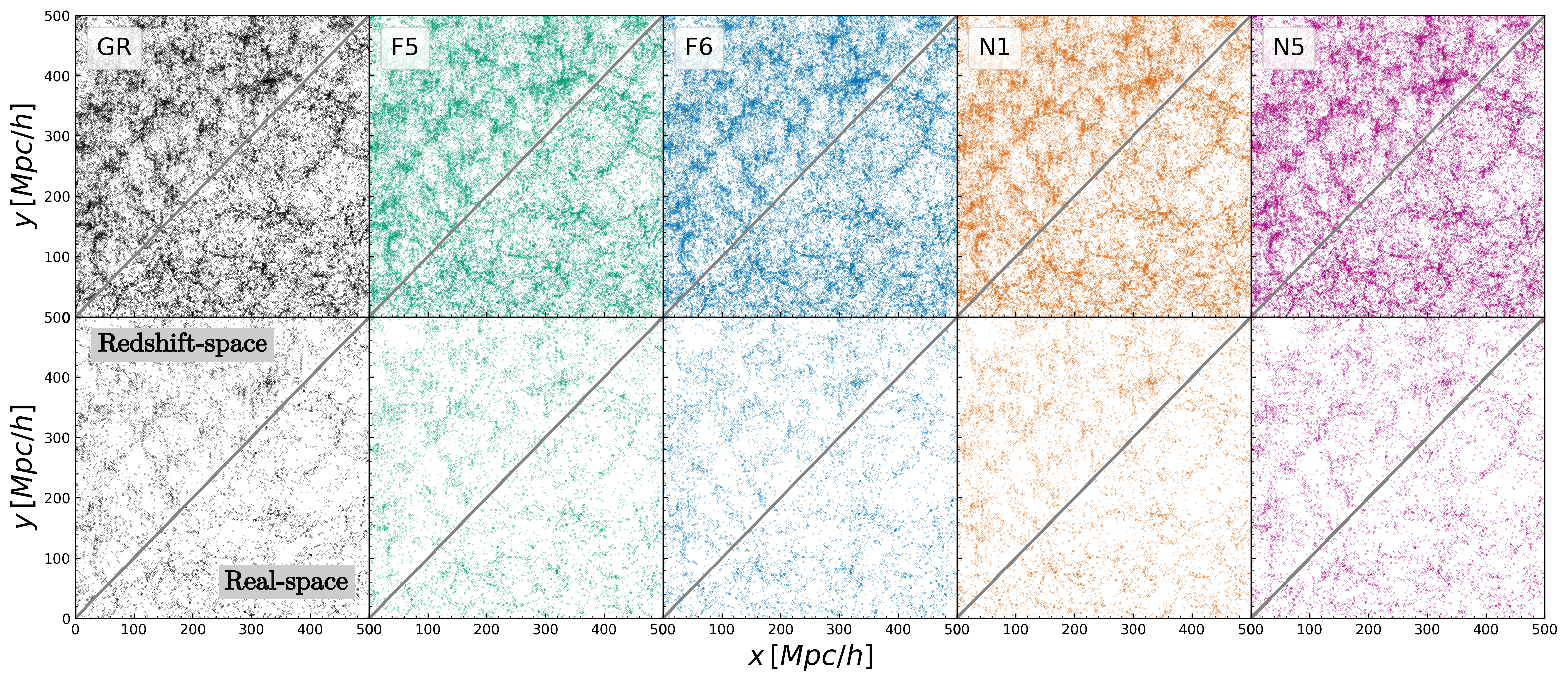}
 \caption{The spatial distribution of DM halos (top panels) and HOD galaxies (bottom panels)
 in a 100\Mpch\ thick slice from the \elephant\ simulations
 at redshift $z=0$. The data represented above (below) the diagonal of each box correspond
 to the redshift (real) space distribution. DM halos are plotted in
 the mass range $[10^{12},\,2\times10^{15}]\,M_\odot/h$.}
 \label{fig:boxes_Halo_density}
\end{figure*}

Figure \ref{fig:MF_papercum} shows the cumulative halo mass function
(cHMF) measured from the \elephant\ simulations and the lower panel shows the relative difference
between the MG and GR models at redshift $z=0.3$.
We compare our results from the \elephant\ simulations to theoretical HMF predictions from
\cite{Tinker2008} (hereafter T08) for both MG models studied. The T08 HMF was calibrated using a spherical overdensity (SO) algorithm
to identify DM halos in \lcdm\ numerical simulations, thus it is consistent with the approach used in \texttt{ROCKSTAR}
to identify halos \cite{Behroozi2013ApJ}. The T08 HMFs for MG models are not drawn in Fig. \ref{fig:MF_papercum} since they have the same behavior as that of GR but shifted, following the data points of each model. As shown by the trend of the T08 HMF in Fig.~\ref{fig:MF_papercum}, $f(R)$ and nDGP predict more halos than the \lcdm\ model for masses above $10^{12}M_\odot/h$ due to the enhancement of gravity
\cite{SchmidtPhysRevD,Schmidt2009PhRvDa, Lombriser2013wta,Cataneo2016JCAP, Arnalte-Mur2017MNRAS,Garfa_2019mnras, Wright_2019JCAP}.

The cHMF is of particular interest since it allows us to select different halo populations from the simulations by
defining thresholds in halo mass,
i.e. selecting halos with
mass above a certain value $M_{min}$ \citep[for details, see
e.g.][]{Arnalte-Mur2017MNRAS}. For that
we set a fixed number density
$\bar{n}(\mathcal{H})$ for each of our halo samples, and define $M_{min}$ in each gravity model separately in such a way to
match these number densities. We define three halo populations, \HI, \HII,
\HIII, with corresponding number densities
$\bar{n}=10^{-4},\,5\times10^{-4}$ and $10^{-3}\,h^3\mathrm{Mpc}^{-3}$,
respectively.

The galaxy sample, hereafter denoted as $\mathcal{G}$, corresponds to central galaxies inside DM halos whose distribution can be approximated by a step-like function parameterized in terms of the properties of their host halos.
A more detailed description of the HOD modeling can be found in Refs. \cite{Zheng2007zg, DESI_ELEPHANT}. Our galaxy sample has a number density defined by $\bar{n}=3.5\times10^{-4}\,h^3\mathrm{Mpc}^{-3}$, which is close to that of the \HII\ halo sample.

\begin{table}
\centering
\caption{Properties of the halo samples used in this work. In each case, we list the minimum halo mass $M_{min}$ used to obtain the required number density $\bar{n}$ for a given redshift $z$ and gravity model.}
\begin{tabular}{cccrrrrrr}
\hline
                           & Halo                    & $\bar{n}$                          & \multicolumn{5}{c}{$M_{min}\,[10^{13}h^{-1}M_\odot]$}                                                                                                                        \\
$z$                          & population              & $[h^3\text{Mpc}^{-3}]$                 & GR                                                  & F5                         & F6                         & N1                         & N5                         \\ \hline\hline
0    & \multicolumn{1}{c}{$\mathcal{H}_1$} & \multicolumn{1}{c}{$10^{-3}$} & 0.319  & 0.371 & 0.377 & 0.326  & 0.322  \\ 
     & \multicolumn{1}{c}{$\mathcal{H}_2$} & \multicolumn{1}{c}{$5\times10^{-4}$}        & 0.724   & 0.833 & 0.805 & 0.740 & 0.729 \\ 
     & \multicolumn{1}{c}{$\mathcal{H}_3$} & \multicolumn{1}{c}{$10^{-4}$} & 3.42  & 4.05 & 3.51 & 3.56 & 3.46 \\ \hline
0.3 & \multicolumn{1}{c}{$\mathcal{H}_1$} & \multicolumn{1}{c}{$10^{-3}$} & 0.317  & 0.387 & 0.368 & 0.334 & 0.321 \\ 
     & \multicolumn{1}{c}{$\mathcal{H}_2$} & \multicolumn{1}{c}{$5\times10^{-4}$}        & 0.705   & 0.817 & 0.744 & 0.728 & 0.710  \\ 
     & \multicolumn{1}{c}{$\mathcal{H}_3$} & \multicolumn{1}{c}{$10^{-4}$} & 3.02  & 3.55 & 3.06 & 3.18 & 3.06 \\ \hline
0.5 & \multicolumn{1}{c}{$\mathcal{H}_1$} & \multicolumn{1}{c}{$10^{-3}$}  & 0.284  & 0.366 & 0.323 & 0.310 & 0.290  \\ 
     & \multicolumn{1}{c}{$\mathcal{H}_2$} & \multicolumn{1}{c}{$5\times10^{-4}$}        & 0.658   & 0.770  & 0.679  & 0.684  & 0.663  \\ 
     & \multicolumn{1}{c}{$\mathcal{H}_3$} & \multicolumn{1}{c}{$10^{-4}$} & 2.64  & 3.10 & 2.65 & 2.80 & 2.67 \\ \hline
\end{tabular}
\label{tab:halopopulations}
\end{table}

The above number densities have been chosen to be representative of the samples used for cosmological
analyses of galaxy surveys such as
the SDSS \cite{kazin2010}, BOSS \cite{sdss_boss_2017}
and eBOSS \cite{BOSS2020arXiv}.
We also take into account the resolution limits of the simulations to make sure that the shot noise would not
dominate our results.
Table \ref{tab:halopopulations}
shows the values of the minimum halo mass $M_{min}$ for each gravity model used to obtain
the required number density for a given redshift.


\section{Clustering measurements}\label{sec:clustering}

We characterize the spatial distribution of cosmic tracers in the \elephant\ simulations (DM particles, halos and galaxies),
in
both real- and redshift-space with the two-point correlation function (2PCF). We use the publicly available code Correlation
Utilities and Two-point Estimation \citep[CUTE, ][]{alonso2013cute} to compute the 2D 2PCF with the Landy-Szalay \cite{LandySzalay_1993} estimator in the comoving range $1<s$ [\Mpch] $<80$, for $20$ linearly spaced bins with constant separation, $\Delta s = 4$\Mpch, in order to compare with previous measurements performed at similar scales \cite{Arnalte-Mur2017MNRAS,Aguayo2019oxg}.

In the non-linear regime, the clustering is well characterized by a much larger variance of the density fluctuations with respect to the one at large scales. In this regime the distortions due to peculiar velocities of the matter tracers may even exceed the Hubble flow producing a smearing effect known as \emph{Fingers of God} (FoG) \cite{Jackson_FoG_1972}. To model the redshift-space distortions we construct mock catalogs from the simulations using
the distant-observer approximation. The positions in real space, $\mathbf{r}$, are converted into redshift space,
$\mathbf{s}$, after adding the contribution due to the peculiar velocities of the tracers along the
line-of-sight (LOS). In
order to reduce the impact of cosmic variance in the redshift space distortions measurements, we average over
three different LOS ($\hat{\mathbf{x}}$, $\hat{\mathbf{y}}$, $\hat{\mathbf{z}}$), then, the errors on
the 2PCF measurements are estimated by the standard deviation over fifteen measurements, obtained from the 5 realizations mentioned in \S\ref{sec:simulations}. All the models
studied here exhibit the same \lcdm\ background expansion history, therefore
all redshift-distance relations are the same among them.
Fig.~\ref{fig:boxes_Halo_density} shows a comparison
of the spatial distribution of tracers in real- and redshift space for both catalogs of halos and galaxies
in a 100\Mpch\ thick slice.

The information about clustering anisotropies can be analyzed by computing multipole moments \cite{kaiser_1987,Hamilton1992b} and clustering wedges \citep{Kazin_2012_estimators, Sanchez2013} from the full 2D 2PCF.
In particular, we characterize the clustering either with the first two
non-vanishing multipole moments, \ie{} the monopole and the quadrupole, or with two clustering
wedges of the 2PCF.
Considering the expansion only up to the hexadecapole, we can express the 2D 2PCF as
\cite{Hamilton1992b}:
\begin{equation}
 \label{eqn:ximultiexp}
 \xi(s,\mu)=\xi_0(s)L_0(\mu)+\xi_2(s)L_2(\mu)+\xi_4(s)L_4(\mu)\, ,
\end{equation}
with $L_l(\mu)$ being the Legendre polynomials of degree $l$, and the
coefficient of the expansion corresponding to the $l^{th}$ multipole
moment of the 2PCF:
\begin{equation}
 \label{eqn:multipoles}
 \xi_l(s) \equiv \frac{2l +1}{2}\int_{-1}^{+1}\, \mbox{d}\mu\, \xi(s,\mu)
 L_l(\mu)\,.
\end{equation}
Here $\mu\equiv\cos{(|\vec{s}|/s_{\parallel})}$ is the cosine of the angle between the separation vector of the tracer pair and the
LOS direction.

An alternative and complementary measure to the multipole moments are the clustering wedges \cite{Kazin_2012_estimators}, which
correspond to the angle-averaged \xiis over wide bins of $\mu$ such that
\begin{equation}
 \label{eqn:wedges}
 \xi_{w}(s) \equiv \frac{1}{\Delta\mu} \int_{\mu_{1}}^{\mu_{2}} \xi(s,
 \mu) \mathrm{d} \mu\,,
\end{equation}
where $\Delta\mu = \mu_{2}-\mu_{1}$ is the wedge width. We consider two clustering wedges,
that is
the \textit{transverse} wedge, $\xi_\perp(s) \equiv \xi_{1/2}(\mu_{min} = 0,
s)$, and the \textit{radial} (or LOS) wedge, $\xi_\parallel(s) \equiv
\xi_{1/2}(\mu_{min} = 0.5, s)$, computed in the ranges $0\leq\mu<0.5$
and $0.5\leq\mu\leq1$, respectively. In this work, both multipole moments and wedges are measured via numerical integration of the full 2D 2PCF in the plane-parallel approximation according to equations
\eqref{eqn:multipoles} and \eqref{eqn:wedges}, respectively.

The deviations in clustering between GR and MG models can be quantified through the linear distortion parameter $\beta\equiv f(z)/b(z)\approx\Omega_m^{\gamma}(z)/b(z)$ where $\gamma$ is the so-called growth index, $f(z)$ is the linear growth rate and $b(z)$ is the linear bias parameter, which relates the cosmic tracers with the DM density field \cite{peebles1980large,Hamilton1993a,peebles1993principles, Wang1998, Lue2004PhRvD, Lahav1991MNRAS,Linder_2005PhRvD,Jennings_RSD_MG_2012,Song2009}. The linear distortion parameter is expressed in terms of the amplitude of the distortions of the clustering. To obtain $\beta$, one can use as estimator the ratio between the redshift-space and real-space monopole, denoted as
$R(s)$, as well as the ratio between the redshift-space quadrupole and the monopole, denoted as $Q(s)$ \cite{Hatton1997RSD_MG,Okumura2011_RSD_MG,percival2011redshift,Clifton2012,He_RSD_2018}. We note however that these estimators only perform well in the linear regime. These quantities can be written as follows \cite{hamilton_review_1998}:
\begin{eqnarray}
 \label{eqn:xi_ratios_multipoles}
 R(s)&=&\frac{\xi_0(s)}{\xi_0(r)} = 1+ \frac{2\beta}{3} +
 \frac{\beta^2}{5},\\
 Q(s)&=&\frac{\xi_2(s)}{\xi_0(s)-\frac{3}{s^3}\int^s_0ds'\xi_0(s')s'^2}
 = \frac{\frac43\beta+\frac47\beta^2}{1+ \frac{2\beta}{3} +
 \frac{\beta^2}{5}}\,,
 \label{eqn:xi_ratios_multipoles2}
\end{eqnarray}
where $\xi_0$ and $\xi_2$ are the monopole and
quadrupole of the 2PCF, respectively.


\section{Results}\label{sec:results}

We now present the results of our analysis, accompanied by a detailed discussion. A busy reader
can skip to the conclusions in \S\ref{sec:discussion_conclusions}, where we cover all the important findings supplemented
by the final discussion. The clustering analysis of the different halo populations and the galaxy sample has been performed at three different cosmic times as mentioned in \S\ref{sec:clustering}. However, below we will focus on $z=0.3$, which is compatible with the BOSS LOWZ Galaxy Sample ($z\lesssim0.4$), and we will only refer to the other redshifts where relevant.

\subsection{Clustering in real space}\label{subsec:real_space}

\begin{figure}
 \includegraphics[width=\linewidth]{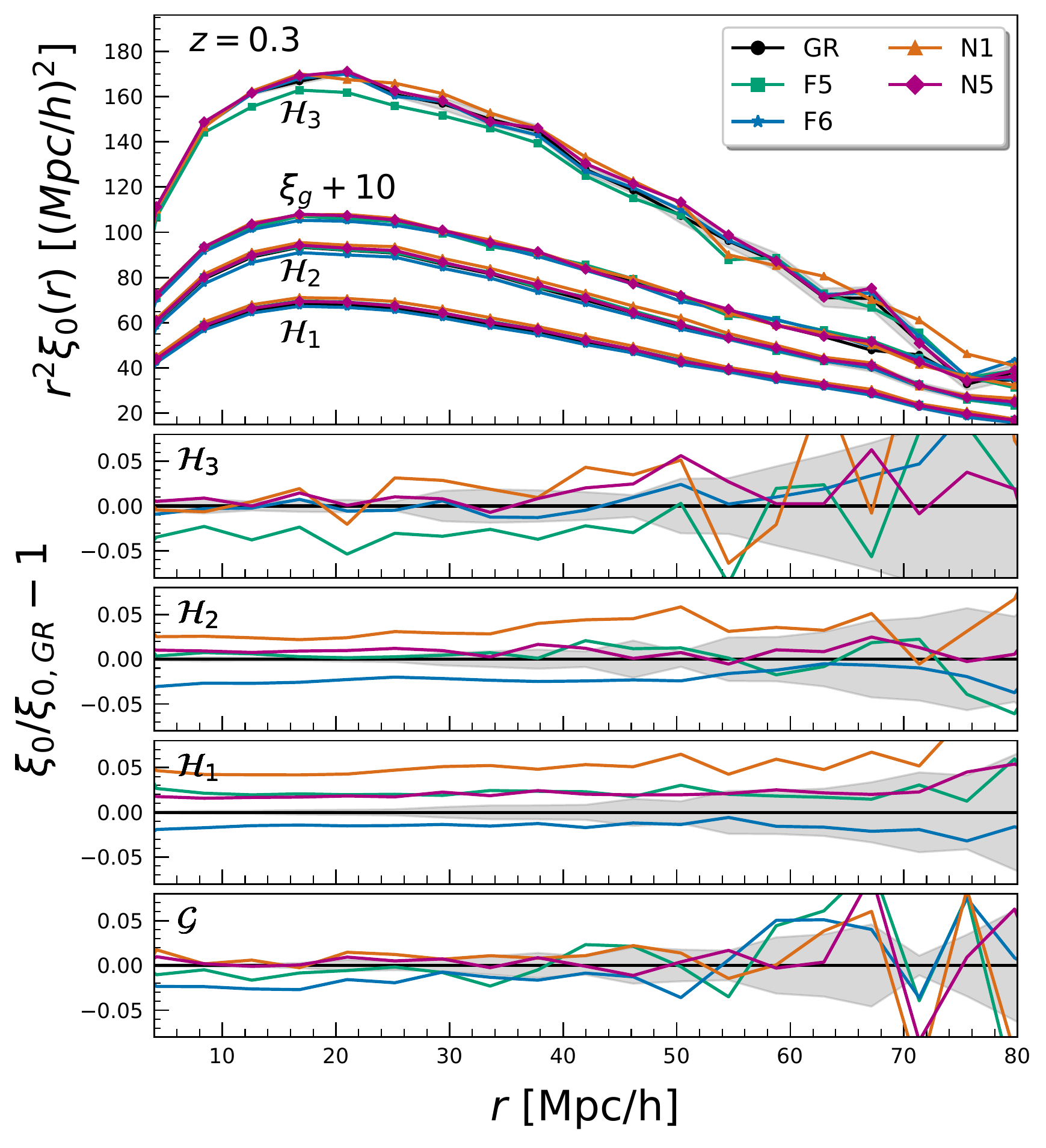}
 \caption{The real-space monopole of the 2PCF of halos and galaxies at
 redshift $z=0.3$, for tracers indicated by the labels. The lower panels show the relative difference of MG models with respect to \lcdm. The grey-shaded areas indicate the propagated measurement errors for \lcdm\ over fifteen measurements.}
 \label{fig:measure_monopole_RS}
\end{figure}

The distribution of matter and halos in the real space is characterized by statistical isotropy among all spatial directions, in a sense of an ensemble average. This means that all the higher multipole
moments of the 2PCF with $l>0$ vanish. Thus, the matter clustering is encoded in the monopole moment of the 2PCF, $\xi_0(r)$. Fig.~\ref{fig:measure_monopole_RS} shows the real-space 2PCF of \elephant\ at redshift $z=0.3$ for the three different halo populations and the galaxy sample, for the considered gravity models. The galaxy correlation function closely follows the result for the \HII\ population, so it was shifted upwards
by 10 units to facilitate a better visualization.
The lower panels show the relative difference of MG models with respect to \lcdm. The results for other redshifts are qualitatively similar, and we do not show them here in the interest of space. Quantitatively, by comparing the 2PCFs, we observe that, for a given MG model, the amplitude of the clustering increases with cosmic time, as expected, being higher for lower the redshifts.
The clustering amplitude depends strongly on the mass-cut of the tracer distribution, and as a consequence, denser samples are much more sensitive to clustering, the extreme case being the one given by the sample \HI.
This variation in the clustering is present in all the MG models investigated, being also an expected feature since the models share
the background history and therefore reproduce a hierarchical structure formation starting from the same initial conditions \cite{Brax88,KoyamaPhysRevD79,Lombriser2015axa}.

For the most sparse sample of \HIII\ all MG models except F5 exhibit the shape and amplitude
consistent with that of the GR case. For the higher number density samples differences among models start to become
more apparent and significant. For \HII, the N1-case experiences enhanced clustering, and this signal appears to be significant
on nearly all scales up to 70\Mpch. This deviation attains a maximum at $r\simeq50$\Mpch, where it amounts to $5\%$ increase of the
$\xi_0$ amplitude. In contrast, the F6 model is characterized by a lower correlation strength over the separation range
of $5\leq r[$\Mpch$]\leq 50$. The N5 and F5 models for this sample exhibit deviations from GR that are typically
smaller than $1\%$. Moving to our highest density halo sample, \HI, we observe very similar trends as for the other
halo populations, but now the N1 and F6 models also foster small, but significant, deviations from the GR case.
The increased statistical significance of the departures that we can appreciate for this sample show a clear advantage
of the increased number density, providing better sampling of the density field. For N5 and F5, the increase of the correlation
amplitude over \lcdm\ is small, at the $2\%$-level, but appears to be significant up to pair separations of $r\simeq 45$\Mpch.

Last but not least, for the galaxy sample, $\mathcal{G}$, the general impression is that
the MG-induced deviations are very much suppressed for all the models. This is not surprising, accounting for the fact
that the HOD mock galaxy catalogs were constructed to have the same (consistent with observations) amplitude of the projected correlation function,
$w_p(r_p)$, up to maximum $1\%$ variation \cite{Cautun2018MNRAS,DESI_ELEPHANT}. Despite this, for F5 and F6 at $r\leq30$\Mpch\ we can still observe some remaining minute differences, below $2\%$.
This is related to the fact that significant differences prevail in the real-space 3D correlation function, suggesting that the chameleon mechanism (operating efficiently on small non-linear scales), leaves a lasting imprint on the $f(R)$ galaxy clustering on those scales \citep[see also][]{Hellwing2013}.

The scalaron fifth force affects more significantly the growth of structures in the F5 model than in F6, that itself mostly agrees with the \lcdm\ clustering signal, especially for low number density samples such as \HIII. This effect is neither appreciable for higher number densities as in \HI\ nor in the galaxy sample as shown in Fig.~\ref{fig:measure_monopole_RS}.
This behavior is well understood considering the weak modifications introduced to GR by the scalar field in the F6 model, therefore only deviations within a few percent
are expected. The above results are in agreement with the analyses performed with different simulations that also include the same MG models \cite{Arnalte-Mur2017MNRAS,Aguayo2019oxg,Wright_2019JCAP, Garfa_2019mnras}.

\begin{figure}
 \includegraphics[width=\linewidth]{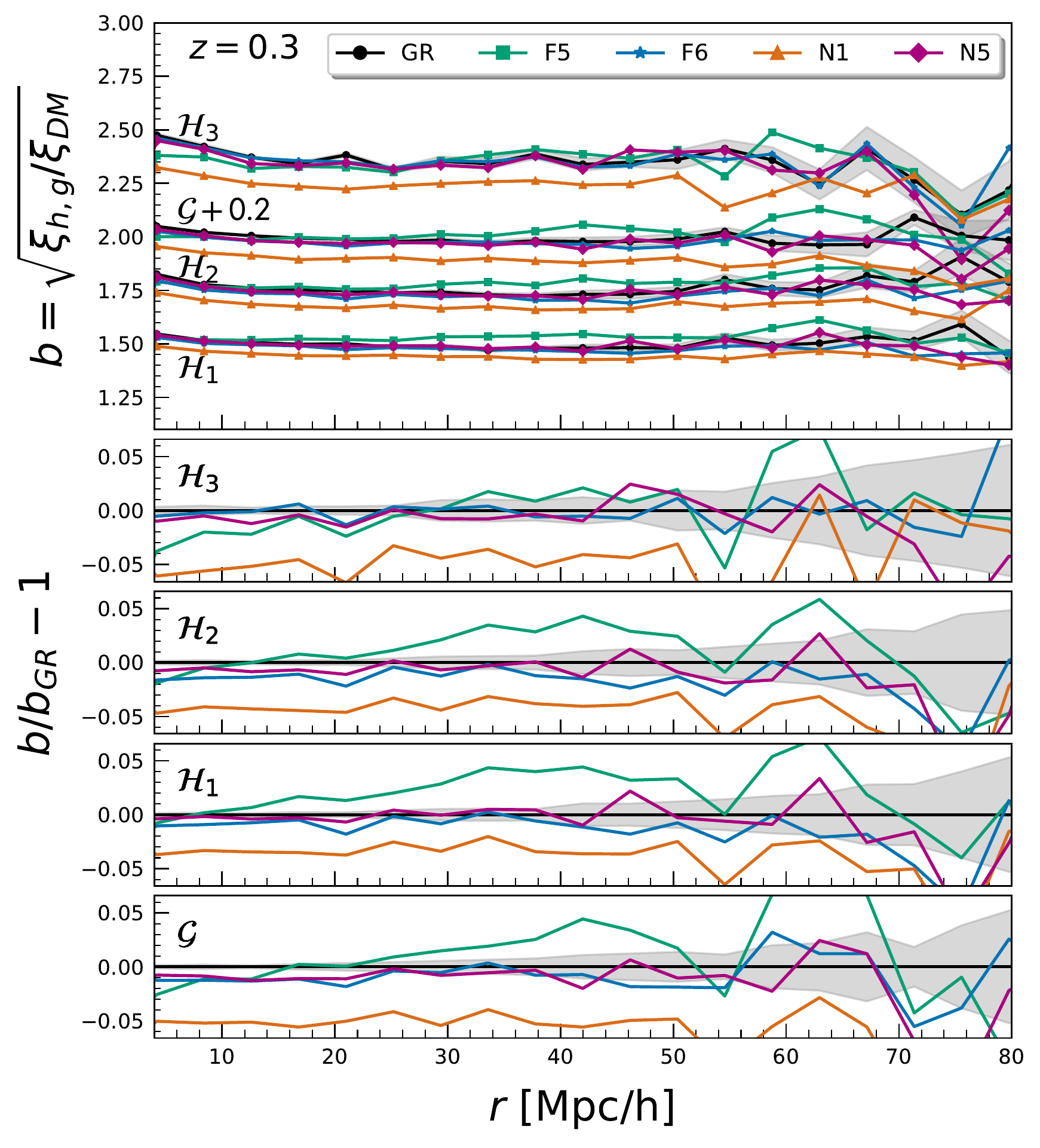}
 \caption{The effective linear halo and galaxy bias as a function of the
 comoving scale at redshift $z=0.3$, for tracers indicated by the
 labels. The lower panels show the
 relative difference of MG models with respect to the \lcdm\
 model. The gray-shaded areas indicate the propagated measurement errors for \lcdm\ over fifteen measurements.}
 \label{fig:bias_HP_G}
\end{figure}

\subsection{Linear halo and galaxy bias}\label{subsec:linbias}

The simplest way of expressing the relation between biased tracers (halos and galaxies), and the underlying smooth DM density field is given by the linear bias parameter $b$. This is defined as $b \equiv \delta_{x}/\delta $, where $\delta$ is the DM density contrast, $\delta_{x}$ is the density contrast
of the tracers, and the subscript $x=\{\mathrm{g},\,\mathrm{h}\}$ denotes galaxies or halos, respectively.
The linear bias can be estimated from the ratio
of the real-space auto-correlation functions of the tracer and DM field as follows:
\begin{equation}\label{eqn:linbias}
b(M,r,z)=\sqrt{\frac{\xi_x(M,r,z)}{\xi_{\mathrm{DM}}(r,z)}}.
\end{equation}
From the above we can expect that
the relation between the density field of tracers and the smooth background can be generalized to a more complicated form, usually involving some scale-dependency. In the following analysis we will not attempt to model any such beyond-linear bias dependence. We are motivated by the fact that the scale-dependence of the bias is weak on the large enough
scales (\ie{} $\geq30$\Mpch) \cite{Smith_2007PhRvD_bias,Basilakos2000bj_bias,Basilakos2011jx_bias,Hoffmann2016_bias,Desjacques2018}.

\begin{table*}
\caption{The effective linear galaxy and halo bias, $b(z)$, estimated from 
Eq.~\eqref{eqn:linbias} for all gravity models at three different redshifts 
$z = 0,~0.3$ and $0.5$.}
\centering
\begin{tabular}{ccm{2cm}m{2cm}m{2cm}m{2cm}m{2cm}m{2cm}}
\hline
$z$                          & Sample              & \centering GR                                                  & \centering F5                         & \centering F6                         & \centering N1                         & \multicolumn{1}{c}{N5}  \\ \hline\hline
0    & \multicolumn{1}{c}{$\mathcal{G}$}   & $1.59\pm0.04$ &	$1.61\pm0.05$ &	$1.61\pm0.05$ &	$1.50\pm0.04$ &	$1.55\pm0.04$ \\
     & \multicolumn{1}{c}{$\mathcal{H}_1$} & $1.30\pm0.03$ &	$1.33\pm0.04$ &	$1.30\pm0.04$ &	$1.27\pm0.03$ &	$1.29\pm0.03$ \\
     & \multicolumn{1}{c}{$\mathcal{H}_2$} & $1.50\pm0.04$ &	$1.52\pm0.05$ &	$1.50\pm0.05$ &	$1.45\pm0.04$ &	$1.48\pm0.04$ \\
     & \multicolumn{1}{c}{$\mathcal{H}_3$} & $2.00\pm0.07$ &	$2.02\pm0.08$ &	$2.02\pm0.08$ &	$1.91\pm0.06$ &	$1.94\pm0.07$ \\ \hline
0.3 & \multicolumn{1}{c}{$\mathcal{G}$}    & $1.80\pm0.05$ &	$1.83\pm0.05$ &	$1.77\pm0.05$ &	$1.67\pm0.05$ &	$1.76\pm0.05$ \\
    & \multicolumn{1}{c}{$\mathcal{H}_1$}  & $1.49\pm0.04$ &	$1.54\pm0.04$ &	$1.47\pm0.03$ &	$1.44\pm0.04$ &	$1.49\pm0.04$ \\
     & \multicolumn{1}{c}{$\mathcal{H}_2$} & $1.75\pm0.05$ &	$1.80\pm0.05$ &	$1.72\pm0.04$ &	$1.68\pm0.05$ &	$1.74\pm0.05$ \\
     & \multicolumn{1}{c}{$\mathcal{H}_3$} & $2.35\pm0.08$ &	$2.35\pm0.08$ &	$2.34\pm0.08$ &	$2.23\pm0.08$ &	$2.34\pm0.08$ \\ \hline
0.5 & \multicolumn{1}{c}{$\mathcal{G}$}    & $1.94\pm0.06$ &	$1.95\pm0.07$ &	$1.95\pm0.06$ &	$1.85\pm0.05$ &	$1.88\pm0.07$ \\
    & \multicolumn{1}{c}{$\mathcal{H}_1$}  & $1.61\pm0.05$ &	$1.63\pm0.05$ &	$1.60\pm0.04$ &	$1.56\pm0.04$ &	$1.58\pm0.05$ \\
     & \multicolumn{1}{c}{$\mathcal{H}_2$} & $1.92\pm0.06$ &	$1.93\pm0.06$ &	$1.91\pm0.05$ &	$1.84\pm0.05$ &	$1.86\pm0.06$ \\
     & \multicolumn{1}{c}{$\mathcal{H}_3$} & $2.58\pm0.11$ &	$2.52\pm0.12$ &	$2.59\pm0.10$ &	$2.47\pm0.10$ &	$2.51\pm0.10$ \\ \hline
\end{tabular}
\label{tab:biasestimated}
\end{table*}

In Fig.~\ref{fig:bias_HP_G} we show the measured halo (galaxy) bias as a function of the comoving scale.
In each case the shaded area corresponds to the mean value and
1$\sigma$ scatter for the \lcdm\ model.
The bottom panels illustrate the relative difference for each sample and each model with respect
to the \lcdm\ case.
The data shown here support our stipulation of very weak scale-dependence of the linear bias, which is confirmed
for all our samples and models.
The trend shown in Fig.~\ref{fig:bias_HP_G} indicates that the linear
bias decreases monotonically with
growing number density for all the models. Moreover, the deviation with respect to the \lcdm\ model is more significant in high density
halo populations, such as \HI\ than in the low density ones, for all models except N1, which exhibits a prominent deviation in the \HIII\ case.
At the same time, the relative deviation in the linear bias is almost constant in all the MG models for scales up to 60\Mpch.

Looking more closely at model-specific cases, it is clear that for both F6 and N5 the differences from the \lcdm{} are minute,
and hence insignificant given our sample variance. This is expected, considering that these two variants
should exhibit the weakest deviations from the standard structure formation scenario.
Consequently, the stronger variants of $f(R)$ and nDGP models are characterized by much clearer and more significant
departures from the GR case. This is especially highlighted for the bias of the N1 model, which takes systematically lower
values then GR for the all the
probed scales. These differences appear relatively flat with distance and amount steadily to about $\sim-5\%$
for all the samples, including galaxies. The F5 variant appears here as the most variable and interesting one: its linear bias difference with respect to the fiducial case exhibits clearly both scale and sample dependence.
For the low-density sample, \HIII, the difference is small, and except for the smallest
separations, non-significant. The situation is quite the opposite for the high density samples, where we observe
a peak of $\sim+4\%$ around $r\sim40$\Mpch. This relative difference appears to be strongly suppressed
at $r\sim55$\Mpch\ (where for \HIII\ it even takes a minus sign), to again grow beyond $+5\%$
at $r\sim60$\Mpch. However, we believe that the feature at $60$\Mpch\ is rather artificial than due to
any genuine physical effects. This is related to the fact that the linear bias is a ratio of two estimators that,
due to the limited-volume
effects become more and more noisy with the growing scale, while the relative difference with respect to the \lcdm{}
case is itself based on a ratio. In addition, we know that all the involved numbers are small. These all
affect significantly the ratio in question, making it very noisy for $r\geq50$\Mpch.

From the above analysis illustrated in Fig.~\ref{fig:bias_HP_G} a very important picture emerges.
The bias of galaxy and halo fields in MG can be significantly different from the GR-case, and the effect
in general can take both the positive and negative sign. Moreover, the differences depend on the
sample density (hence implicitly on the tracer mass). This has been already pointed out by
some earlier work \cite{samplingbiasSutter2013ssy, biasLazeyras2016xfh}.
The sample density could potentially introduce a degeneracy of the MG effect with the growth rate, therefore the latter parameter could be much closer to the GR expectation than it really is.

\begin{figure}
 \includegraphics[width=\linewidth]{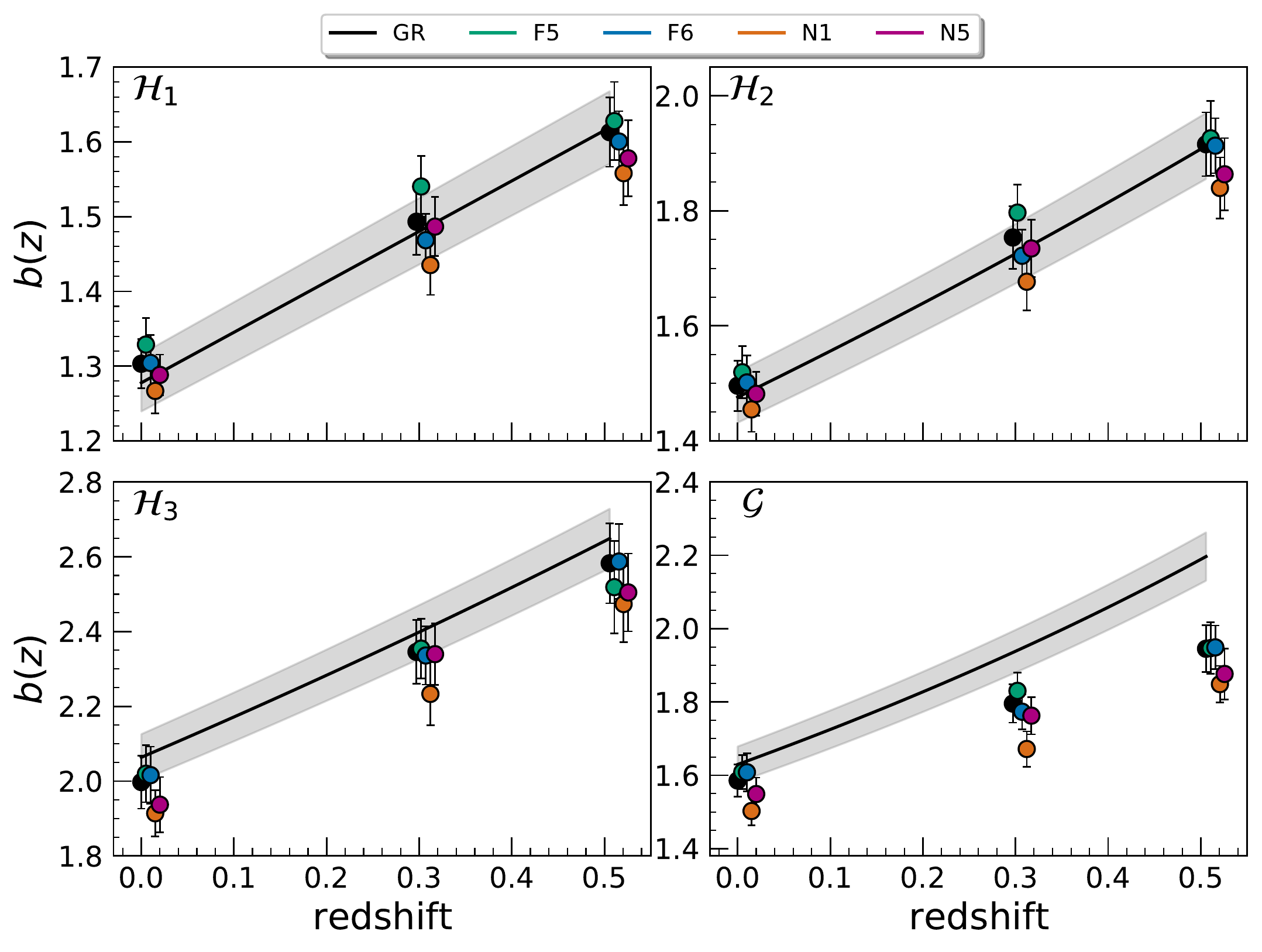}
 \caption{The effective halo (galaxy) bias averaged in the range $40 <$ r[\Mpch] $<80$ as a function of redshift for
 the models considered in this work. The plot shows the bias at redshifts $z=0,~0.3$ and $0.5$, however, for a better
 visualization the redshift of the MG models has been shifted from the mean value. Solid line show the theoretical
 \lcdm\ prediction computed according to the Tinker et. al. \cite{Tinker_bias_2010} formula.
 The error bars represent
 the propagated statistical noise and the gray-shaded areas show a $3\%$ error.}
 \label{fig:biasz_plot}
\end{figure}

Following eq. \eqref{eqn:linbias}, we estimate the effective linear bias $b$ based on a $\chi^2$ fit to the square root of the ratio of the tracer 2PCF and the DM 2PCF
over the comoving range $40<r[$\Mpch$]<80$ for each model, sample and redshift from \elephant\ simulations.
Table
\ref{tab:biasestimated} shows a summary of the measured galaxy and halo bias
for all the considered gravity models at three
different redshifts $z= 0,\,0.3$ and $0.5$. Fig.~\ref{fig:biasz_plot} displays the evolution of the effective bias as
a function of redshift with error bars that represent the $1\sigma$ statistical error.
The solid lines correspond to the theoretical prediction
computed using the Tinker et. al. \cite{Tinker_bias_2010} formula and by averaging the linear bias, $b(M, z)$, of the selected sample as follows:
\begin{equation}
 \label{eqn:biaseff}
 b_{\rm eff}(z) = \frac{\int_{M_{\rm min}}^{M_{\rm max}}n(M, z) b(M,
 z) d M}{\int_{M_{\rm min}}^{M_{\rm max}} n(M, z) dM},
\end{equation}
where the mass limits $[M_{\rm min}$,~$M_{\rm max}]$ have been defined
in Section \ref{sec:simulations}, while $n(M, z)$ and $b(M, z)$, are estimated using the
Tinker et. al. \cite{Tinker2008} mass function and the Tinker et. al. \cite{Tinker_bias_2010} bias model,
respectively. The trend in the bias confirms that the nDGP models differ the most from the standard model predictions both for halo and galaxy samples. Instead, the $f(R)$ models are in good agreement with \lcdm, and they deviate less at low redshifts than the nDGP ones.

\begin{figure}
 \includegraphics[width=\linewidth]{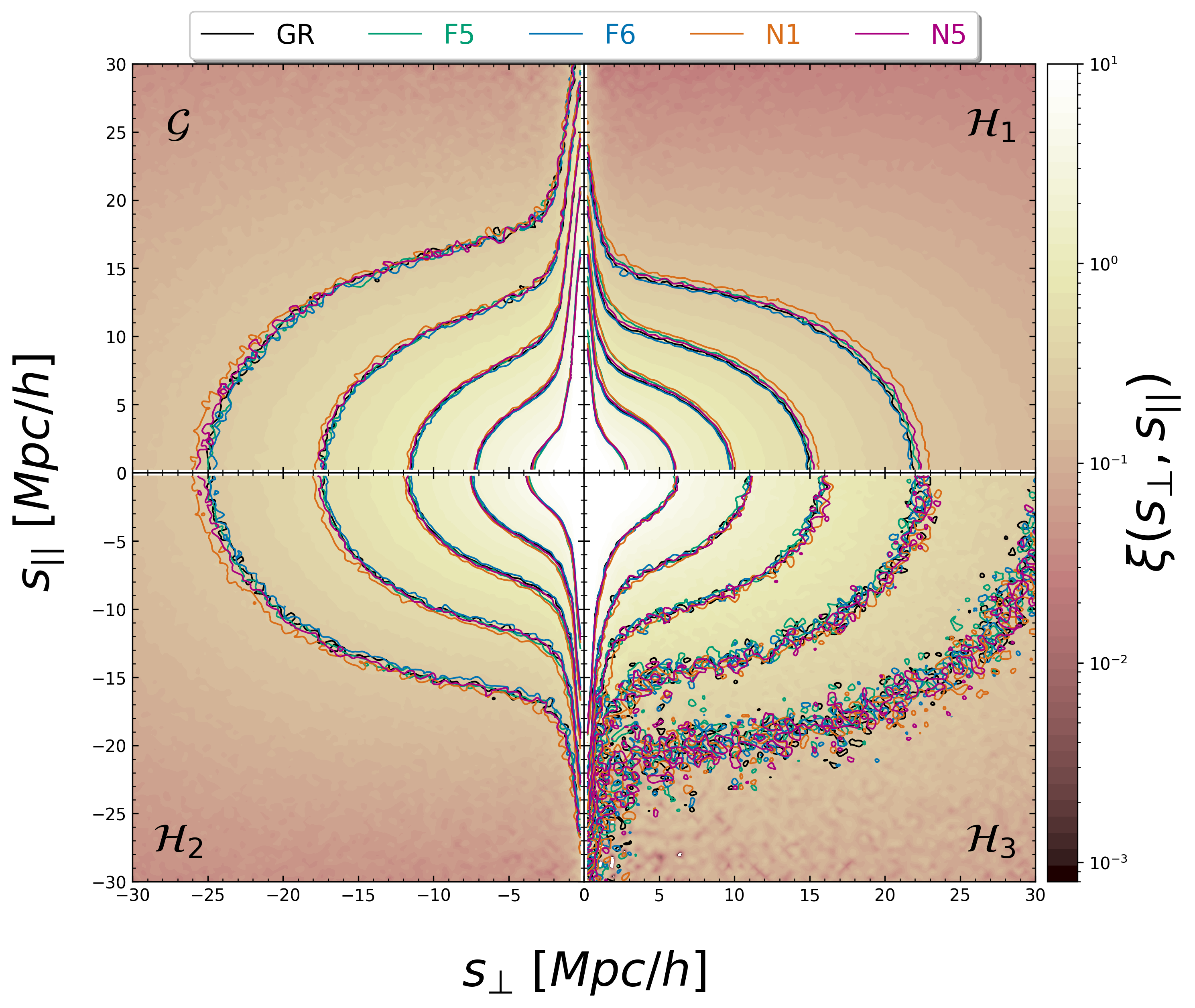}
 \caption{Iso-correlation contours of $\xi(s_\perp, s_\parallel)$ at
 the reference redshift $z=0.3$ for the models indicated in the legend, shown for
 correlation amplitude levels \xiis=$\{0.25, ~0.5, ~1, ~2, ~5\}$. Each quadrant corresponds to a different tracer sample (halos and galaxies) as labeled. The color bar to the right and the background color of the panels indicate the amplitude of \xiis for \lcdm\ as the reference model.
 }
 \label{fig:HPxi-2D}
\end{figure}


\begin{figure*}
 \includegraphics[width=0.45\linewidth]{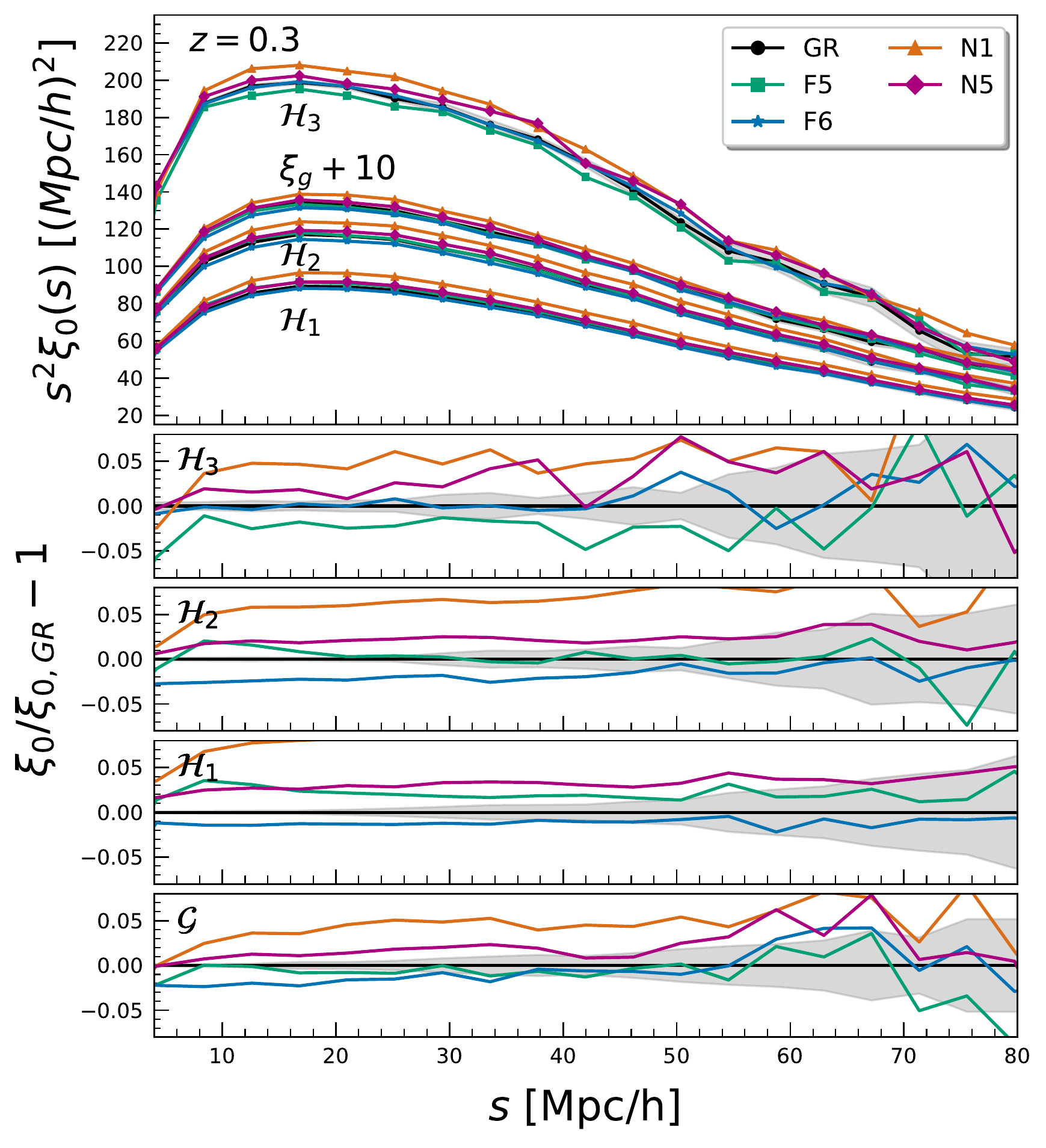}
 \includegraphics[width=0.45\linewidth]{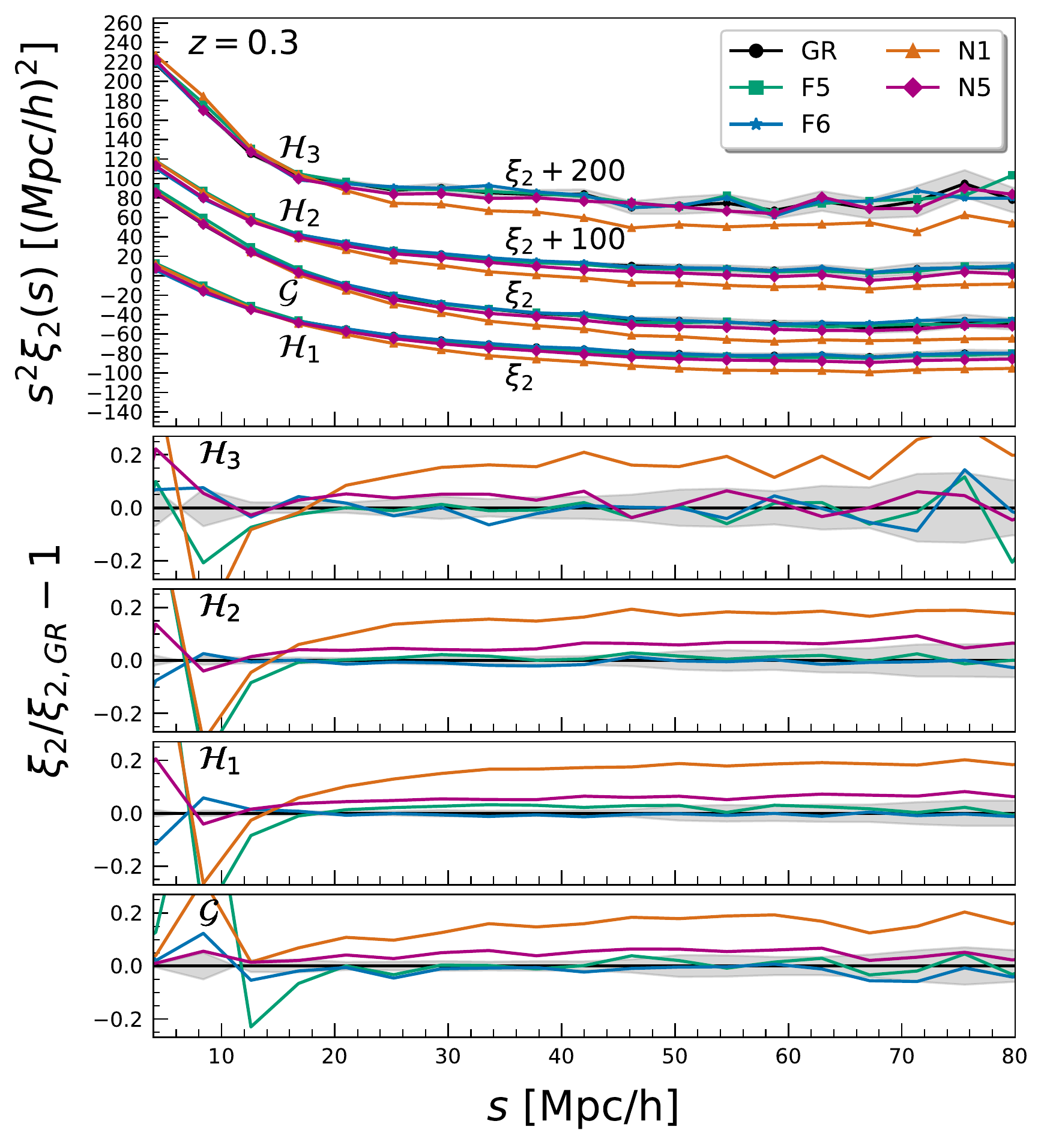}
 \includegraphics[width=0.45\linewidth]{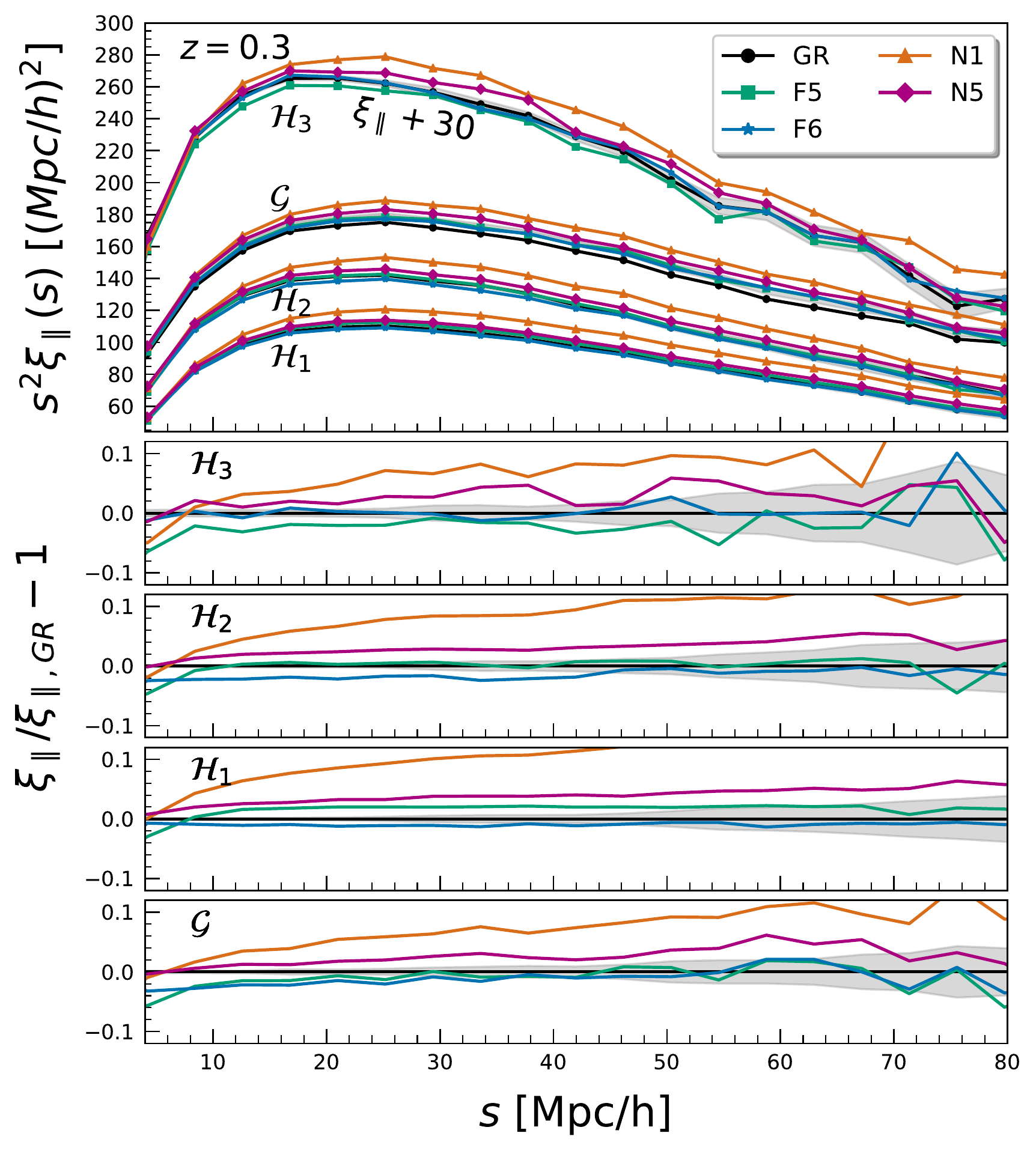}
 \includegraphics[width=0.45\linewidth]{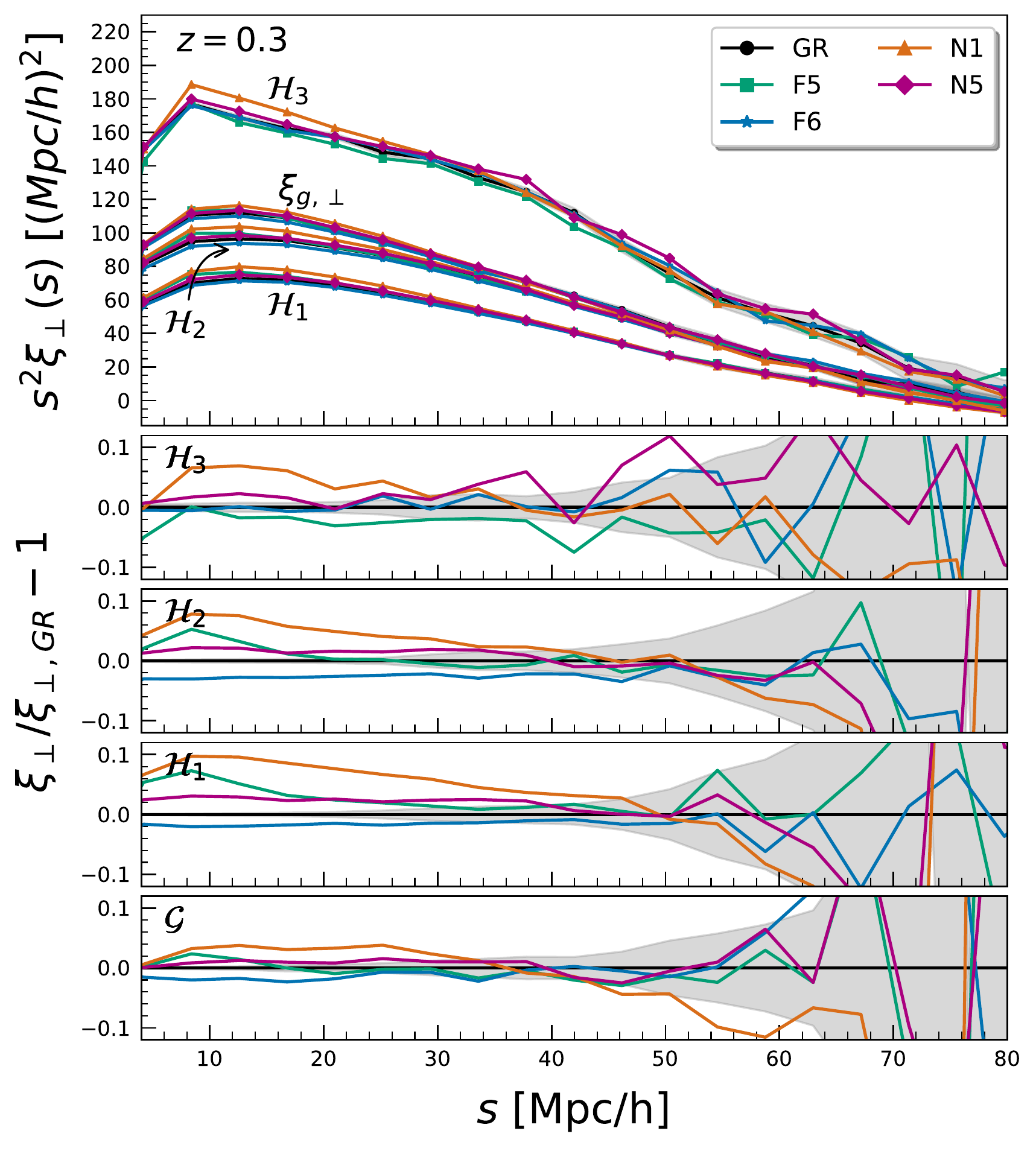}
 \caption{The redshift-space multipole moments of the 2PCF, $\xi_0$ and $\xi_2$ (upper panels), and clustering wedges $\xi_\parallel$ and $\xi_\perp$ (lower panels), for the different gravity models and tracer samples used in this work as indicated by the labels. The lower subpanels in each row show the difference of MG models with respect to the \lcdm\ measurements. The gray-shaded areas correspond to the standard deviation for \lcdm\ over fifteen measurements obtained from 5 different realizations.}
 \label{fig:measure_multipoles_ZS}
\end{figure*}

\subsection{Clustering in redshift space}
\label{subsec:redshift_space}

Figure \ref{fig:HPxi-2D} shows the full 2D 2PCF computed at a redshift $z=0.3$ as a function of the
transverse, $s_\perp$, and parallel, $s_\parallel$, separations with respect to the LOS.
The black line marks the result for the \lcdm\ model from \elephant\ simulations, while each MG model is depicted by a different color. Each quadrant shows the 2PCF signal for a different tracer sample (galaxies and halo populations), as labeled. The iso-correlation contours of \xiis are measured in the amplitude range $[0.25,~5]$.
The resulting contours of galaxy samples for the different gravity models are very close to each other, with the slight exception of the N1 model.
This is not surprising, as the correlation in redshift space is determined, to the first order,
by the amplitude of real-space correlations \cite{kaiser_1987}.
The latter are by design set to be equal among all the mock catalogs. A more careful statistical inspection of the moments of 2D 2PCF would be needed to highlight MG effects for our galaxy samples.

Halo clustering, likewise to mock galaxies, shows slight differences for all MG models. These are best visible for the populations with medium, \HII, and high number densities, \HI. The clustering signals from the \HIII\ populations are noise-dominated at all scales, as expected, and therefore there is no visible distinction between the MG models and GR.
In the most abundant halo population, \HI, there is a clear difference between the GR case and the N1 model which displays an excess in the clustering.
On the contrary, the $f(R)$ gravity variants exhibit a slight suppression in the clustering signal at all scales.
This behavior can be also seen, although to a lesser extent, at small comoving scales, $\sim 4$\Mpch, despite a very similar amplitude in the FoG for all MG models. However, a conclusive analysis on these scales would require high-resolution simulations plus the implementation of additional physical processes which are relevant at those scales.
In the three halo selections for all our MG models, the 2PCF for N1
deviates the most from the GR case.
For the other models the differences, albeit noticeable in the 2PCF iso-contours, are much harder to depict.
In general, as the halo number density decreases,
the differences between MG and GR get smaller.
In addition, the selection effects related to the changing number density (which itself is here driven by the minimum mass cut) shows how RSD are affected at all scales. The amplitude of clustering varies more in the transverse direction than in the LOS, indicating that the Kaiser squashing effect is more sensitive to the coherent motion of the tracers than FoG under the same mass selection.

The complicated pattern and relatively small differences among the models, as seen in the 2D 2PCF, suggest that a more elaborate analysis is needed to robustly quantify MG effects.
As previously pointed out in Sec.~\ref{sec:clustering}, projecting $\xi(s_\perp,s_\parallel)$ onto
one-dimensional statistics (multipole moments or wedges) increases the clustering signal and disentangles the directional dependence on clustering. In Fig.~\ref{fig:measure_multipoles_ZS} we show the redshift space
multipole moments, $\xi_l(s)$ (upper group of plots),
and clustering wedges, $\xi_w(s)$ (lower group of plots),
measured for our halo and galaxy samples at $z = 0.3$ for the different MG models as labeled.
In each case the lower subpanels show the relative differences with respect to the \lcdm\ measurements.

As far as the monopole, $\xi_0(s)$, is concerned, we firstly observe that it decreases as the number density increases mainly due to the difference in the bias
of each sample, as expected. The clustering signal of the medium number density halo sample, \HII, follows closely the monopole of the galaxy sample, which in Fig.~\ref{fig:measure_multipoles_ZS} has been shifted by 10 units for visual convenience.
For all our samples
both nDGP models exhibit a rise in the correlation amplitude
seen at all probed scales.
There is also a general trend of
the brane-world gravity enhanced clustering signal that is dropping with decreasing
sample number density.
In contrast, the situation for the $f(R)$-gravity family is quite different. Here, the effect on the monopole amplitude
goes generally in the opposite direction, and the clustering at fixed number density seems to be slightly suppressed.
The exception from this is F5 in the richest halo sample of \HI. Another observation regarding the correlation function monopole is that the differences in this statistic between the MG models and the fiducial \lcdm\ case generally do not correlate with the magnitude of deviations (e.g. level of the fifth force) of the particular MG variants.
While for nDGP the monopole signal for N1 is always larger than for N5, in the chameleon $f(R)$ for some samples (\ie{}
$\mathcal{G}$, and \HII) it is F6 variant that departs more prominently, while for others (\ie{}
\HI, and \HIII) the F5 deviates more. Finally, while for the brane-world models
the excess clustering signal is relatively flat with the pair-separation scale, in $f(R)$ the differences
are scale-dependent and suppressed for $s\geq 60$\Mpch.

Overall, the behavior of $\xi_0(s)$ that we described above supports the known picture in which the $f(R)$ gravity models are characterized by a very high degree
of non-linearity, even at larger scales. This is a clear virtue of the chameleon screening mechanism: its effectiveness is a function of the local environment, so it operates in a more complicated manner than the Vainshtein screening of nDGP.
The non-linear and non-monotonic behavior of the $f(R)$ clustering signal, as compared to nDGP, could be in principle
exploited for differentiating these two classes of screening mechanisms in the data.

\begin{table*}[htbp]
\caption{The the best-fitting values of the linear distortion parameter, $\beta(z)$, of halos and galaxies, obtained using the $Q(s)$ estimator, Eq. \eqref{eqn:xi_ratios_multipoles2}.}
\begin{center}
\begin{tabular}{ccm{2.5cm}m{2.5cm}m{2.5cm}m{2.5cm}m{2.5cm}m{2.5cm}}
\hline
$z$ & Sample & \centering GR & \centering F5 & \centering F6 & \centering N1 & \multicolumn{1}{c}{N5} \\ \hline\hline
\multicolumn{1}{c}{0} & $\mathcal{G}$ & $0.279\pm0.017$ & $0.283\pm0.018$ & $0.279\pm0.019$ & $0.321\pm0.021$ & $0.289\pm0.017$ \\ 
 & $\mathcal{H}_1$ & $0.349\pm0.017$ & $0.352\pm0.016$ & $0.351\pm0.018$ & $0.385\pm0.016$ & $0.361\pm0.017$ \\ 
 & $\mathcal{H}_2$ & $0.304\pm0.020$ & $0.307\pm0.018$ & $0.305\pm0.019$ & $0.340\pm0.018$ & $0.316\pm0.018$ \\ 
 & $\mathcal{H}_3$ & $0.234\pm0.028$ & $0.233\pm0.027$ & $0.231\pm0.031$ & $0.266\pm0.029$ & $0.239\pm0.031$ \\ \hline
\multicolumn{1}{c}{0.3} & $\mathcal{G}$ & $0.328\pm0.019$ & $0.334\pm0.018$ & $0.330\pm0.019$ & $0.377\pm0.019$ & $0.342\pm0.018$ \\ 
 & $\mathcal{H}_1$ & $0.405\pm0.016$ & $0.403\pm0.016$ & $0.407\pm0.017$ & $0.449\pm0.016$ & $0.422\pm0.019$ \\ 
 & $\mathcal{H}_2$ & $0.346\pm0.019$ & $0.351\pm0.020$ & $0.354\pm0.020$ & $0.391\pm0.019$ & $0.363\pm0.018$ \\ 
 & $\mathcal{H}_3$ & $0.266\pm0.030$ & $0.264\pm0.024$ & $0.266\pm0.030$ & $0.305\pm0.026$ & $0.265\pm0.029$ \\ \hline
\multicolumn{1}{c}{0.5} & $\mathcal{G}$ & $0.340\pm0.020$ & $0.345\pm0.021$ & $0.342\pm0.020$ & $0.388\pm0.022$ & $0.353\pm0.021$ \\
 & $\mathcal{H}_1$ & $0.419\pm0.017$ & $0.421\pm0.016$ & $0.426\pm0.018$ & $0.466\pm0.019$ & $0.427\pm0.017$ \\ 
 & $\mathcal{H}_2$ & $0.357\pm0.018$ & $0.360\pm0.020$ & $0.359\pm0.017$ & $0.403\pm0.020$ & $0.363\pm0.018$ \\ 
 & $\mathcal{H}_3$ & $0.272\pm0.030$ & $0.276\pm0.028$ & $0.275\pm0.025$ & $0.302\pm0.034$ & $0.282\pm0.028$ \\ \hline
\end{tabular}
\end{center}
\label{tab:betameasurements}
\end{table*}

\begin{figure}
 \includegraphics[width=\linewidth]{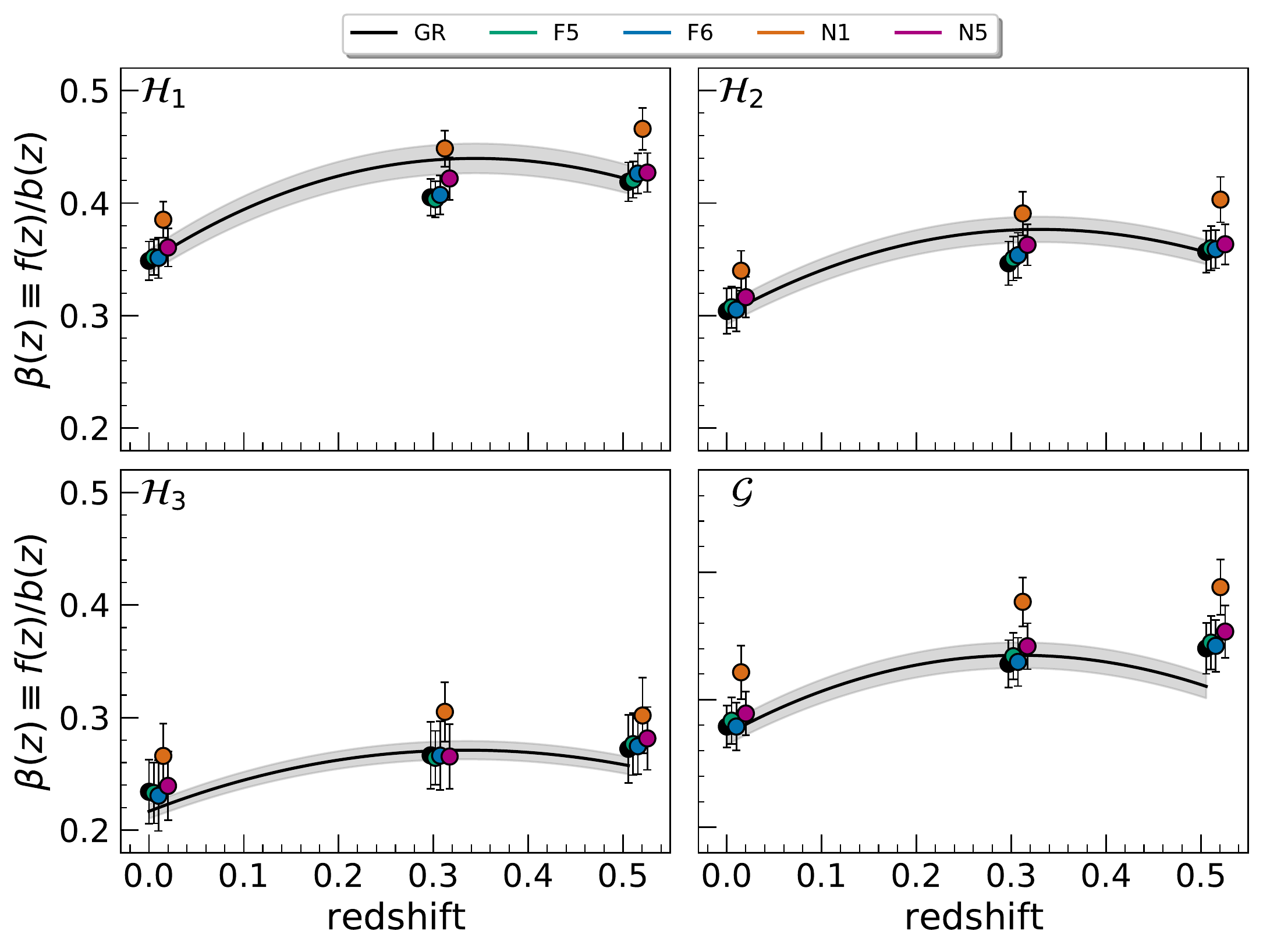}
 \caption{The linear distortion parameter, $\beta(z)$, of halos (galaxies) from \elephant\ simulations as a function
 of redshift for the models considered in this work. For better visualization the datapoints for the MG models have
 been shifted horizontally from the mean values $z=0,0.3$, and $0.5$. Solid lines show the theoretical \lcdm{} prediction computed
 according to the Tinker et. al. \cite{Tinker_bias_2010} bias prescription and growth rate of fiducial cosmology. The error bars represent the propagated statistical noise and
 the gray shaded areas show a 3\% error on $\beta$ for comparison.}
 \label{fig:beta_plot}
\end{figure}

The quadrupole moment of the 2PCF, $\xi_2(s)$, illustrated in the upper-right panel of Fig.~\ref{fig:measure_multipoles_ZS}, encodes the degree of anisotropy generated by RSD. For all our samples,
there is only one model, N1, which fosters clear deviations from the fiducial GR case, while
the effect for N5 is only marginally significant. Taking into account the scatter, both $f(R)$ models exhibit a signal that is statistically consistent with the GR case. The noisy behavior that we see at $s\leq20$\Mpch, rather than a real physical feature, is likely a manifestation of sampling noise to which the quadrupole moment is more susceptible.
The noise at those small pair separation is additionally increased owing to the fact that we consider only halos and central galaxies, and due to the halo-exclusion effect the pair number counts are suppressed at small scales. There could be, in principle, a real physical MG signal encoded at those small scales, but the properties of our simulations and catalogs are not designed to grant such analysis at this time.

The analysis of the clustering wedges (see Fig.~\ref{fig:measure_multipoles_ZS}, lower panels) paints a picture complementary to the one offered by the multipoles.
It is worth noting that the clustering anisotropies as
measured in each of two angular wedges disclose
stronger MG signal than it was in the case for the multipole moments.
Here, the relative deviation from GR of the LOS wedge, $\xi_\parallel(s)$,
for the nDGP models
reaches 10\% (5\%) for N1 (N5).
This is roughly a
factor of two more
than we saw in
the monopole analysis.
What is even more important, is that the increased relative difference is not affected
by the statistical error,
which remains at
a level similar to that of
the monopole measurements. Our results agree here with the earlier work of Barreira et. al. \cite{Barreira2016}, who studied clustering wedges
for similar variants of nDGP models. This opens up an exciting
avenue of using the clustering wedge statistics to hunt for the MG signal in galaxy clustering.

The changes in the clustering are also manifested in the transverse wedge, $\xi_\perp(s)$,
which gives valuable information from an observational point
of view considering the degeneracy between RSD and other observational phenomena such as the Alcock-Paczy\'nski effect \cite{AlcockPaczynski1979Natur}.
The difference between GR and MG seen in the amplitude of the transverse wedge reveals a few interesting facts.
First, we notice that
for the perpendicular projection
the deviations in clustering of galaxy samples for all MG models are much smaller than it was in the case for the LOS direction. This is driven by the fact that the same projected clustering, $w_p(r_p)$,
was imposed in
the construction of mock galaxy catalogs.
In contrast, different halo populations reveal somewhat larger deviations from the GR case.
 Second, the MG effect seen in $\xi_\perp(s)$ seems to be contained
only to small scales, $s\leq 40$\Mpch. Here, the maximal deviations are attained at the highly non-linear
separations of 10\Mpch.
Furthermore, the transverse wedge of the N1 model deviates from GR by about 10\% at small scales $s<20$\Mpch, although this behavior is also displayed by the multipole moments.
In this case the angular
clustering signal is well characterized by a monotonically decreasing function, being almost independent of the number
density of the halo sample.

\subsection{Linear distortion parameter $\beta$}

All the effects revealed by our analysis of the multipoles and wedges should lead to measurable differences in the linear-theory growth-rate parameter, $f$,
as estimated by various combinations of the clustering statistics.
Since we are measuring clustering of biased traces (\ie{} halos and galaxies), we cannot directly measure the growth rate. Instead, we can estimate the linear distortion parameter $\beta$ by fitting the combinations of the measured
multipole moments,
$Q(s)$,
to the function given by Eqn. \eqref{eqn:xi_ratios_multipoles2}. Following earlier related studies
\cite[e.g.][]{DESI_ELEPHANT, Aguayo2019oxg}, we only keep the results derived from $Q(s)$, since it already contains
 the information from $R(s)$ in its denominator. The dependence on the quadrupole suggests that the $Q(s)$ estimator could be more prone to noise than the $R(s)$ ratio alone.
However, it is more advantageous to use the former as it does not depend on the shape of $\xi(r)$ \cite{Hamilton1992b,Hawkins_2dF2003MNRAS}. Thus, when comparing with observational data, no assumptions on the DM clustering signal in the real space are required. To estimate $\beta(z)$ from the quadrupole ratio $Q(s)$ we perform a $\chi^2$ minimization for each of the models, samples and redshifts separately.

Figure~\ref{fig:beta_plot} shows our $\beta$ measurements as a function of redshift for all our samples and models.
The $1\sigma$ error bars on $\beta$ have been estimated via direct propagation of the uncertainties of $Q(s)$ since it is a one-parameter model.
The solid line represents the linear theory prediction for
GR computed using the bias prescription of Ref. \cite{Tinker_bias_2010} and growth rate of the fiducial cosmology,
while the shaded region corresponds to a 3\% deviation from it. This percentage is for reference, and it is well below the current errors on $\beta$ obtained from observations such as the 6-degree Field Galaxy Survey (6dFGS).
Adams et. al. \cite{Adamsbeta6dF2020dzw} have recently reported statistical uncertainty on $\beta$ of about 15\% and systematic one of $\sim17\%$ from this survey.
Table \ref{tab:betameasurements} displays the best-fit values of $\beta$ for all the samples and models investigated. As it can be appreciated, our constraint for the linear distortion parameter at $z=0$ is in good agreement with the
value by
\cite{Adamsbeta6dF2020dzw}, $\beta = 0.289^{+0.044}_{-0.043}$(stat)$\pm0.049$(sys), which was based on joint measurements of RSD and peculiar velocities from the 6dFGS.

\begin{figure}
 \centering
 \includegraphics[width=0.95\linewidth]{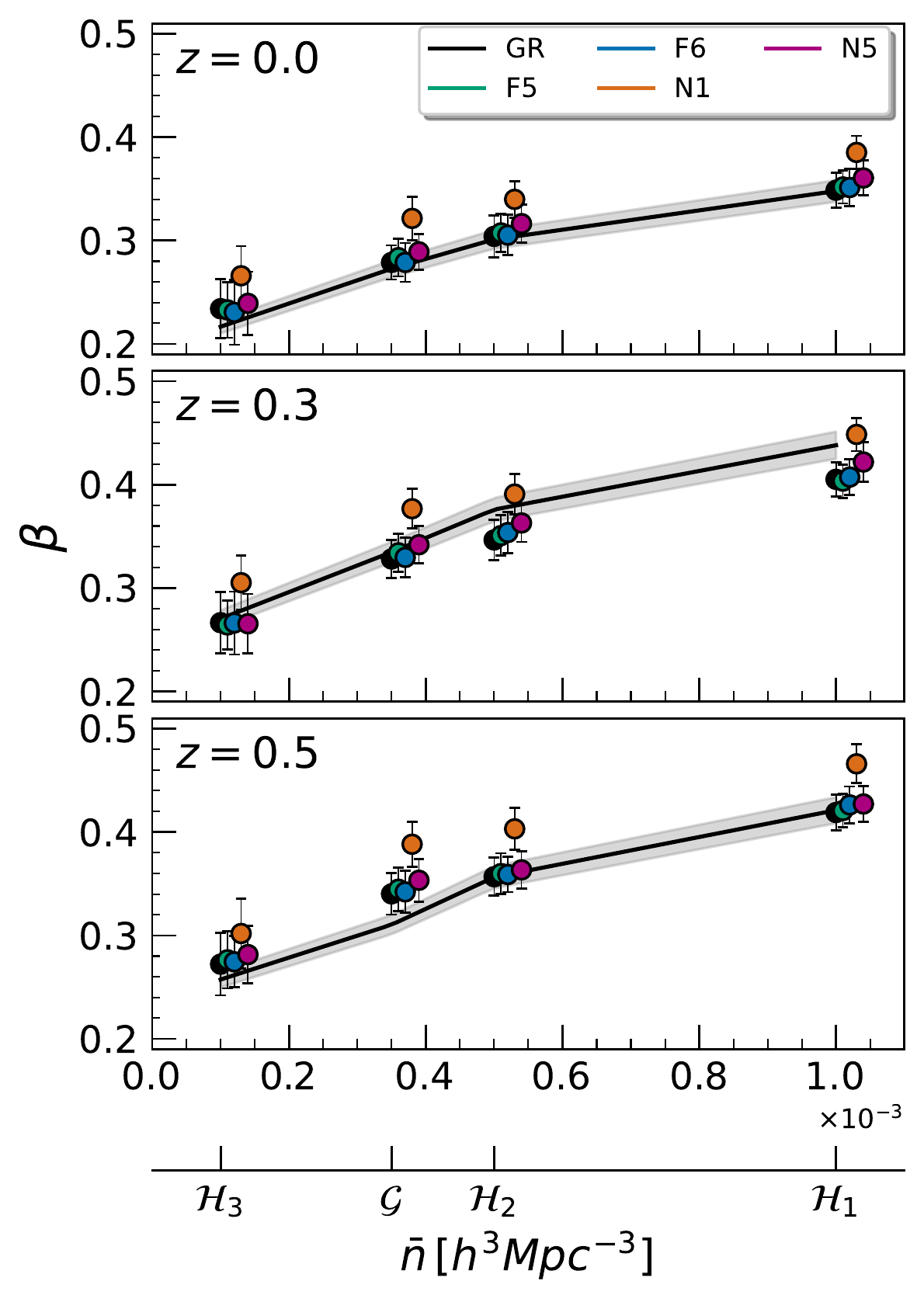}
 \caption{The linear distortion parameter, $\beta(z)$, as a function of the number density of the tracer samples used in this work (halos and galaxies). Results for three different redshifts, $z = 0,\, 0.3$ and $0.5$, are shown as indicated by the labels. The error bars represent the propagated statistical noise and the gray shaded areas show a 3\% error.
}
 \label{fig:beta_densities}
\end{figure}

In Figure \ref{fig:beta_densities} we additionally show the $\beta$ parameter values as a function of the number density of our samples. The results are plotted for three different redshift snapshots, $z = 0,\, 0.3$ and $0.5$, as indicated by the labels.
The solid black line represents the linear-theory prediction for $\beta$ at the given number densities of the samples in the GR scenario. The
points that correspond to particular
number densities are connected using straight lines for visualization.
The shaded region also corresponds to a 3\% deviation from the GR prediction, as in Fig.~\ref{fig:beta_plot}. Beside synthesizing our previous findings, Fig.~\ref{fig:beta_densities} allows us to additionally describe
the behavior of $\beta$ in terms of both the sample density (hence corresponding tracer halo mass) and its redshift. In particular, we note a monotonic increase of $\beta$ with increasing number density. Although the
changes in the predicted slope are not dramatic when comparing the different redshifts, the slight shift of $\beta$ shows its sensitivity with respect to the sample density, and consequently the impact of the degeneracy between the halo or galaxy bias, and the growth factor.

The $Q$ estimator offers in principle
also a possibility to
determine the growth rate. To do so, however, one needs to either obtain information on the value of the sample galaxy
bias parameter or assume one.
In general, it is not always guaranteed that a simple linear theory prediction for the bias will lead to an accurate growth rate estimate from the distortion parameter. In some cases the linear bias relation might not be an adequate description of the data, and therefore higher order bias approximations may be required to account for the scale dependency \cite{Taruya1998stochasticbias, Matsubara1999StochasticityNonlocality, Yoshikawa2001stochasticbiasing, Sato2013stochasticbiasing}.
In fact, the Kaiser \cite{kaiser_1987} enhancement is sensitive to the physics on non-linear scales. Some of these effects might not be captured entirely
by our relatively low resolution N-body simulations. In particular, when $\beta$ is determined from
the quadrupole to monopole ratio, the systematic errors are significant even at the scales considered. This can be well appreciated by noticing that even our \lcdm{}
measurements are not in a perfect agreement with the linear theory predictions.
For this reason, we attempt to recover the true value of the distortion parameter in a bias-independent fashion. As described in the following Sec. \S\ref{sec:RelClustering}, we generalize the clustering estimator initially proposed by Arnalte \etal\ \cite{Arnalte-Mur2017MNRAS}, which allows us to analyze the clustering signal with less impact of halo and galaxy bias and at the same time increase the signal-to-noise ratio of the correlation function.

\begin{table*}[htbp]
\caption{The amplitude of the multipole moments (upper part) and clustering wedges (lower part) of the 2PCF for GR at the reference scale $s_\text{ref}$ used in the clustering ratios.}
\begin{tabular}{ccp{2.5cm}p{2.5cm}p{2.5cm}p{2.5cm}}
\hline
\multicolumn{ 1}{c}{$z$} & \multicolumn{ 1}{c}{Sample} & \multicolumn{ 2}{c}{$\xi_0(s_\text{ref})$} & \multicolumn{ 2}{c}{$\xi_2(s_\text{ref})$} \\ 
\multicolumn{ 1}{c}{} & \multicolumn{ 1}{c}{} & \multicolumn{1}{c}{\footnotesize$16$\Mpch} & \multicolumn{1}{c}{\footnotesize$64$\Mpch} & \multicolumn{1}{c}{\footnotesize$16$\Mpch} & \multicolumn{1}{c}{\footnotesize$64$\Mpch} \\ \hline\hline
0 & $\mathcal{H}_\text{ref}$ & $0.314\pm0.001$ & $0.011\pm0.001$ & $-0.141\pm0.001$ & $-0.018\pm0.001$ \\ 
\multicolumn{1}{c}{} & $\mathcal{G}_\text{ref}$ & $0.449\pm0.001$ & $0.015\pm0.001$ & $-0.109\pm0.002$ & $-0.021\pm0.001$ \\ \hline
0.3 & $\mathcal{H}_\text{ref}$ & $0.317\pm0.001$ & $0.011\pm0.001$ & $-0.165\pm0.001$ & $-0.021\pm0.001$ \\ 
\multicolumn{1}{c}{} & $\mathcal{G}_\text{ref}$ & $0.441\pm0.002$ & $0.0140\pm0.0004$ & $-0.145\pm0.003$ & $-0.024\pm0.001$ \\ \hline
0.5 & $\mathcal{H}_\text{ref}$ & $0.313\pm0.001$ & $0.0100\pm0.0003$ & $-0.174\pm0.001$ & $-0.021\pm0.001$ \\ 
\multicolumn{1}{c}{} & $\mathcal{G}_\text{ref}$ & $0.436\pm0.001$ & $0.014\pm0.001$ & $-0.165\pm0.002$ & $-0.024\pm0.001$ \\ \hline\\ \hline
\multicolumn{ 1}{c}{$z$} & \multicolumn{ 1}{c}{Sample} & \multicolumn{ 2}{c}{$\xi_\perp(s_\text{ref})$} & \multicolumn{ 2}{c}{$\xi_\parallel(s_\text{ref})$} \\ 
\multicolumn{ 1}{c}{} & \multicolumn{ 1}{c}{} & \multicolumn{1}{c}{\footnotesize$16$\Mpch} & \multicolumn{1}{c}{\footnotesize$64$\Mpch} & \multicolumn{1}{c}{\footnotesize$16$\Mpch} & \multicolumn{1}{c}{\footnotesize$64$\Mpch} \\ \hline\hline
0 & $\mathcal{H}_\text{ref}$ & $0.261\pm0.001$ & $0.0040\pm0.0004$ & $0.367\pm0.001$ & $0.0180\pm0.0004$ \\ 
\multicolumn{1}{c}{} & $\mathcal{G}_\text{ref}$ & $0.408\pm0.002$ & $0.490\pm0.001$ & $0.490\pm0.002$ & $0.023\pm0.001$ \\ \hline
0.3 & $\mathcal{H}_\text{ref}$ & $0.255\pm0.001$ & $0.0030\pm0.0004$ & $0.379\pm0.001$ & $0.0190\pm0.0004$ \\ 
\multicolumn{1}{c}{} & $\mathcal{G}_\text{ref}$ & $0.387\pm0.002$ & $0.496\pm0.001$ & $0.496\pm0.002$ & $0.023\pm0.001$ \\ \hline
0.5 & $\mathcal{H}_\text{ref}$ & $0.248\pm0.001$ & $0.0020\pm0.0004$ & $0.378\pm0.001$ & $0.0180\pm0.0004$ \\ 
\multicolumn{1}{c}{} & $\mathcal{G}_\text{ref}$ & $0.374\pm0.002$ & $0.498\pm0.001$ & $0.498\pm0.002$ & $0.023\pm0.001$ \\ \hline
\end{tabular}
\label{tab:srefclustratios}
\end{table*}

\begin{figure*}
 \includegraphics[width=0.45\linewidth]{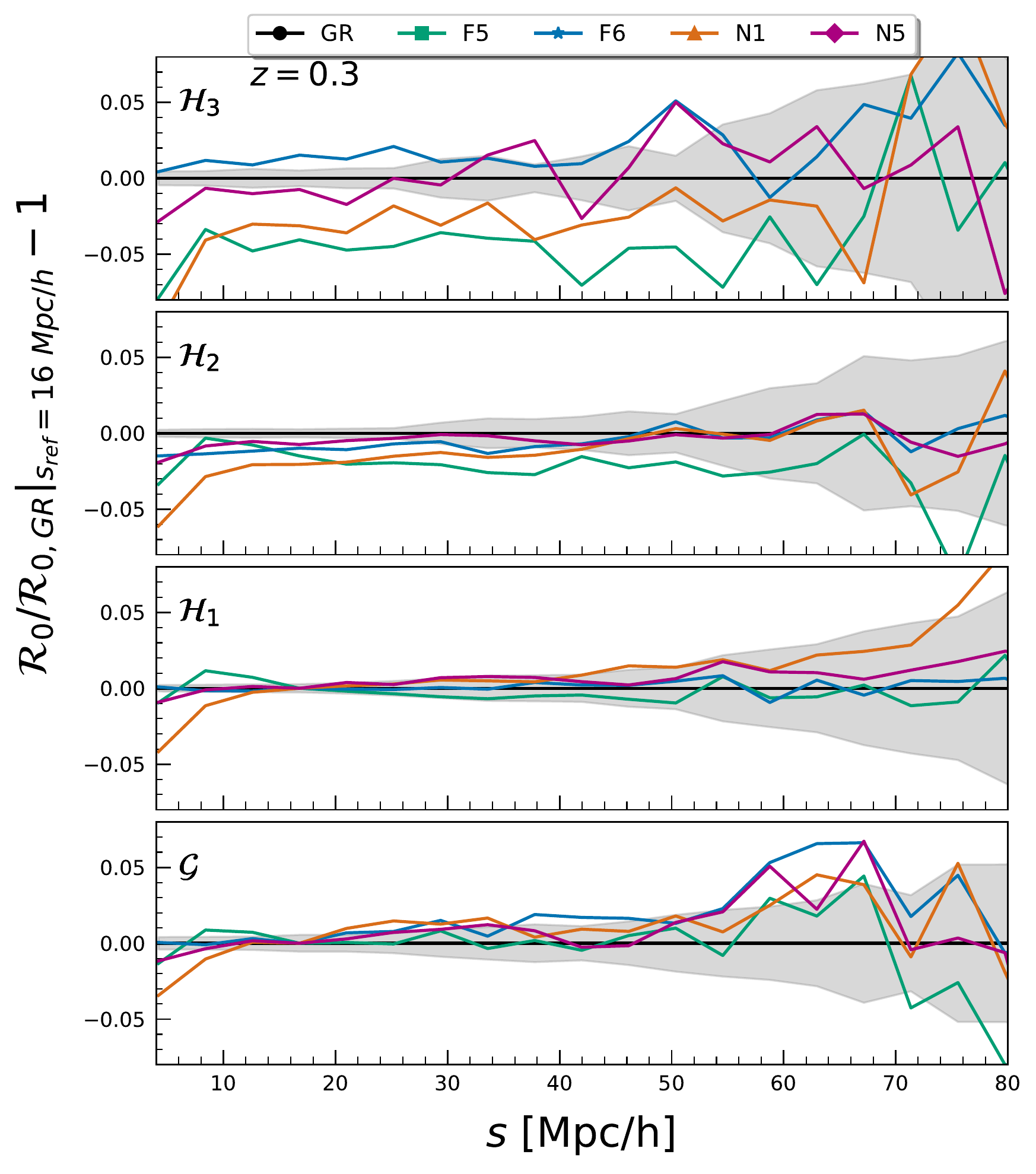}
 \includegraphics[width=0.45\linewidth]{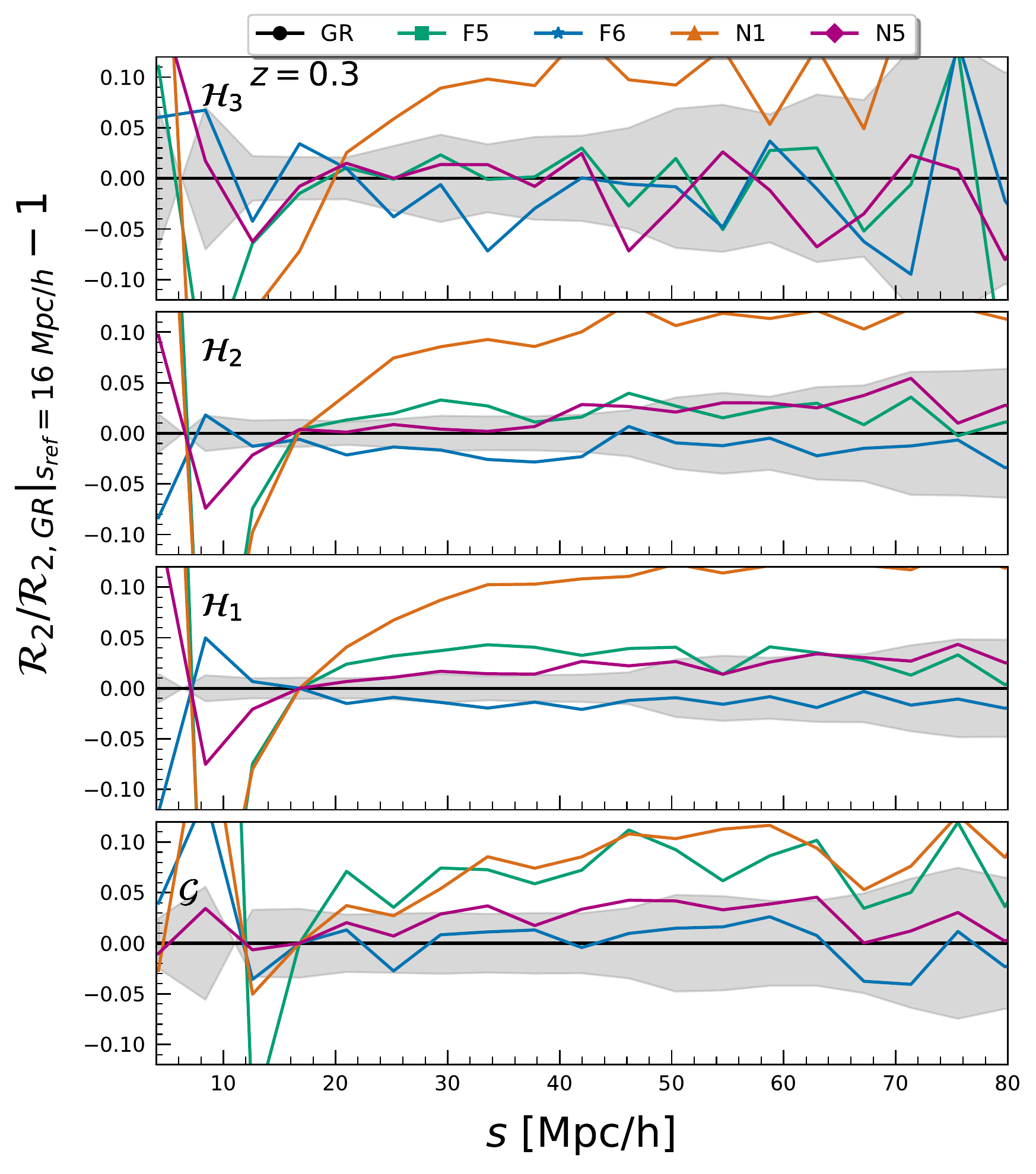}\\
 \includegraphics[width=0.45\linewidth]{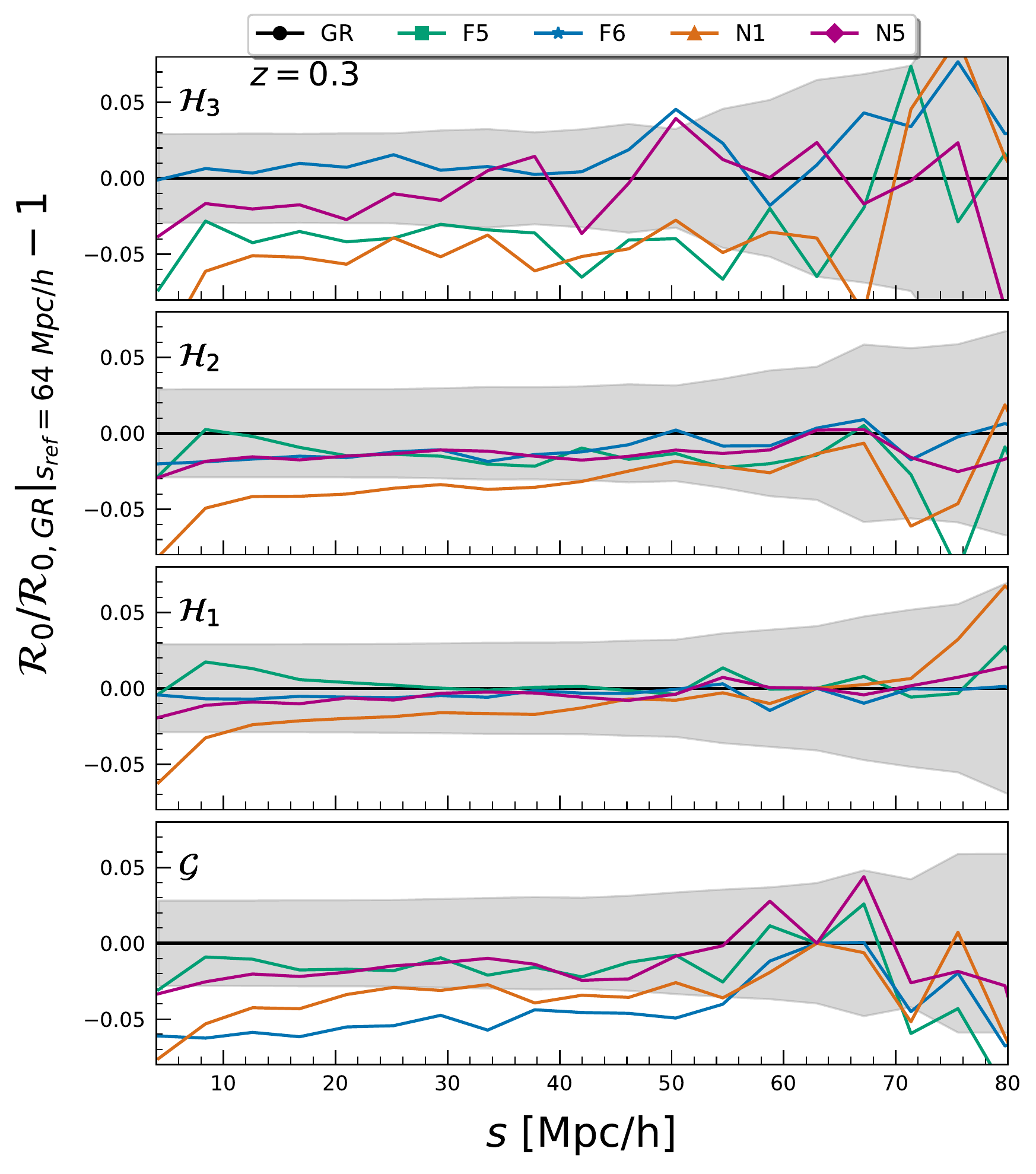}
 \includegraphics[width=0.45\linewidth]{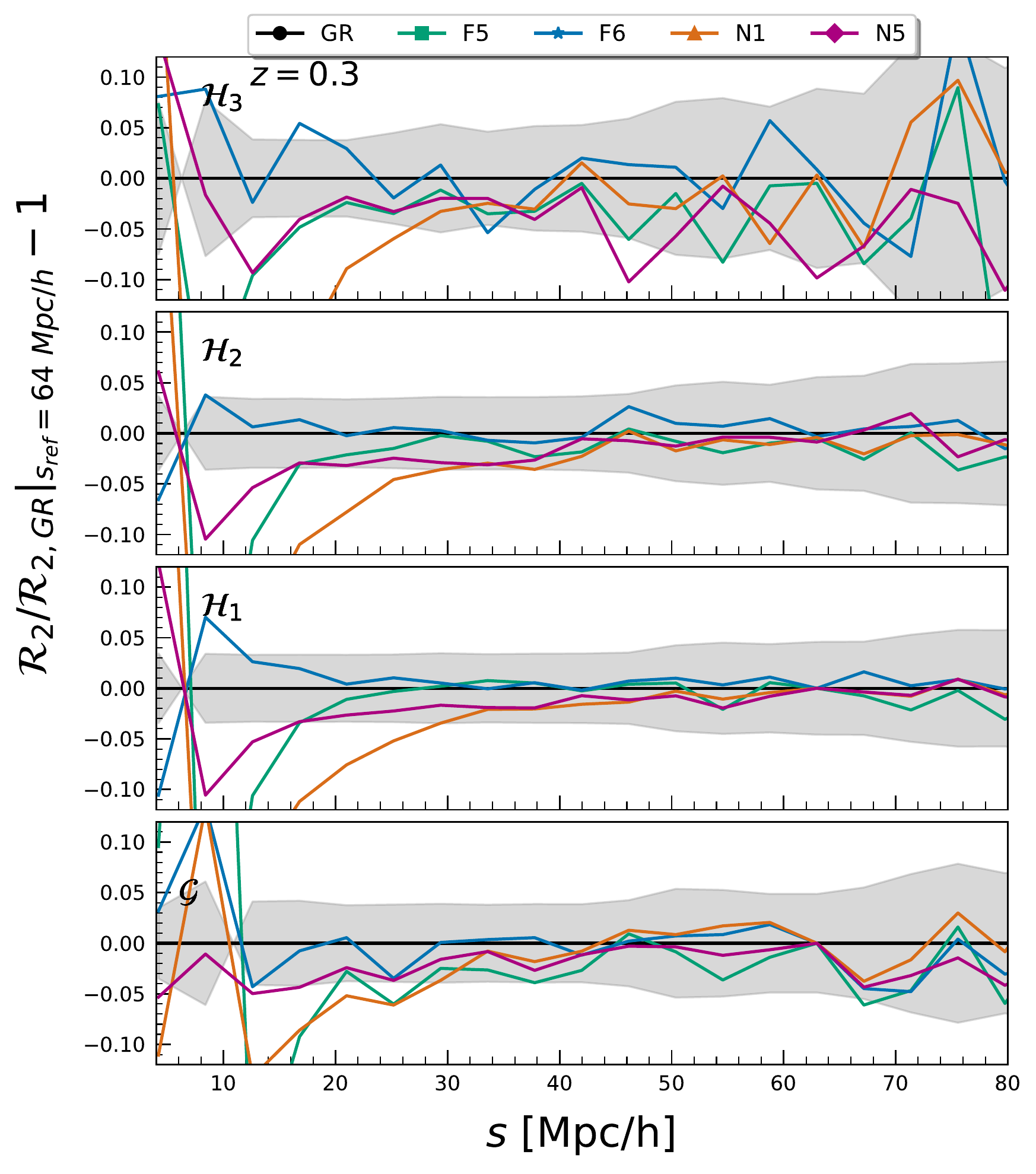}
 \caption{The relative difference of the MG clustering ratios with respect to the \lcdm\ measurements for both the monopole (left) and quadrupole (right) of the 2PCF of halos (galaxies) as indicated by the labels, at two different reference scales $s_\text{ref}=16$\Mpch\ (upper panels) and $s_\text{ref}=16$\Mpch\ (lower panels). The gray-shaded areas correspond to the standard deviation for GR over fifteen measurements of the 2PCF estimation.}
 \label{fig:HP_ZS_RelClust_xi02}
\end{figure*}

\begin{figure*}
 \includegraphics[width=0.45\linewidth]{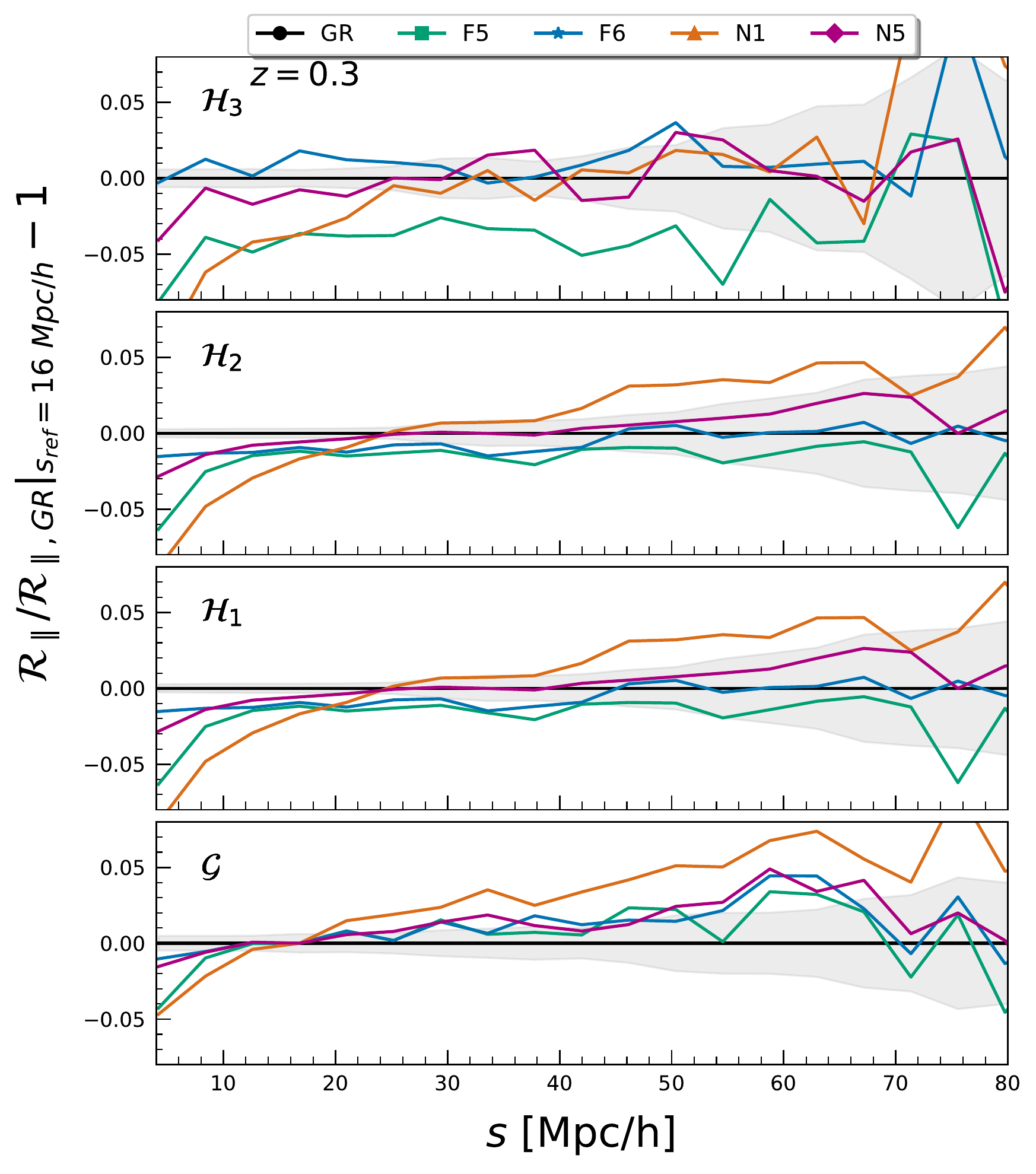}
 \includegraphics[width=0.45\linewidth]{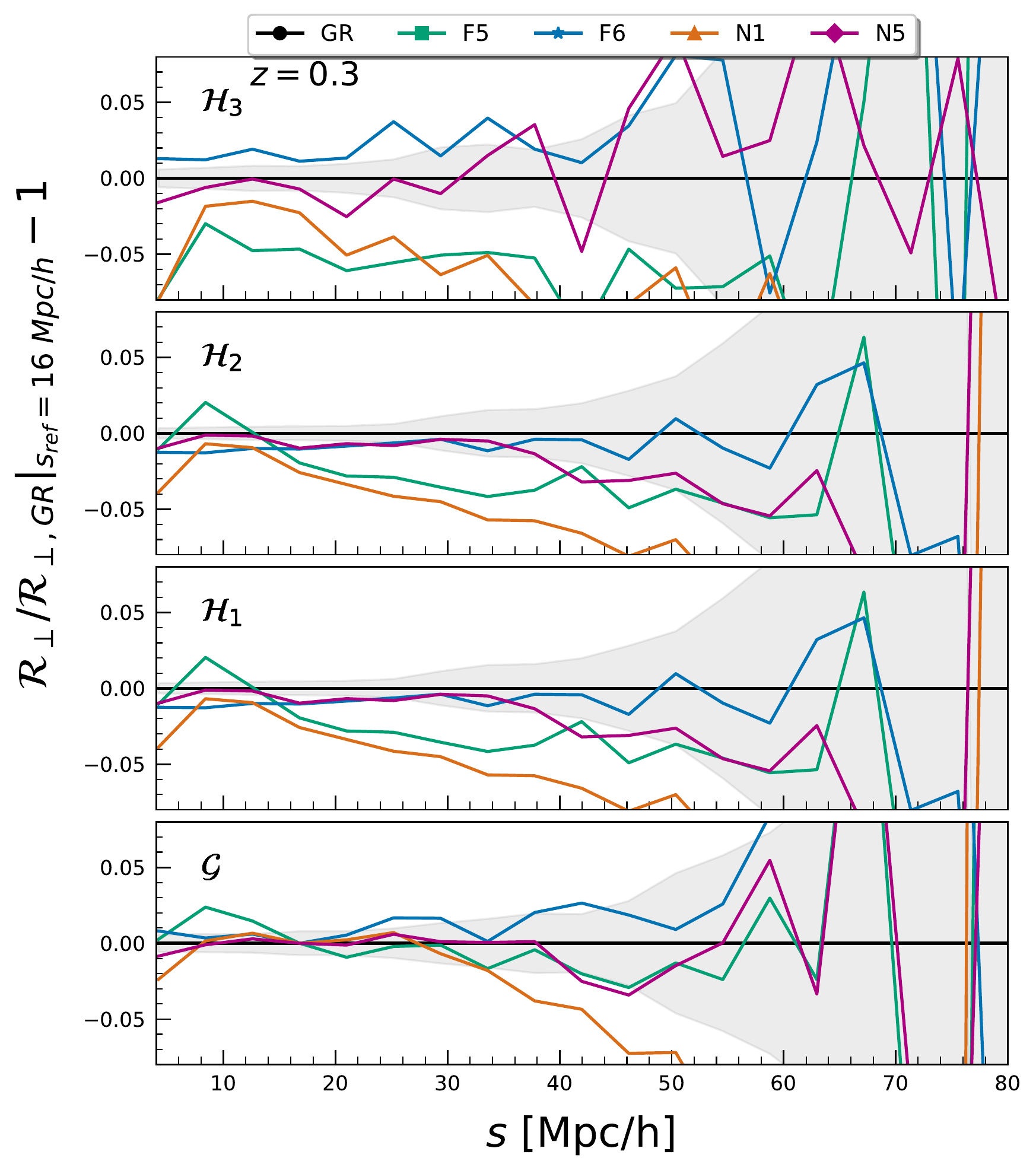}\\
 \includegraphics[width=0.45\linewidth]{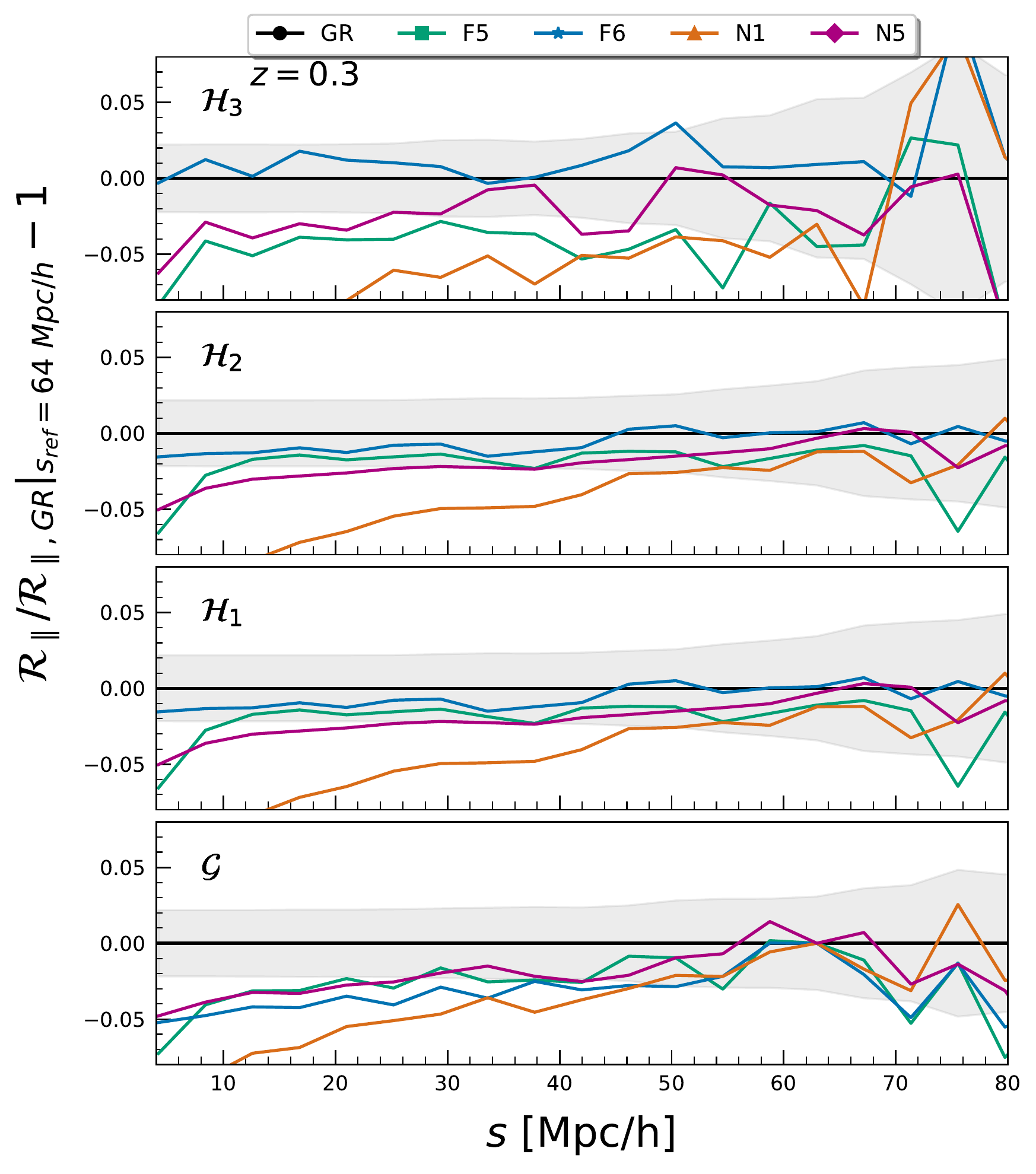}
 \includegraphics[width=0.45\linewidth]{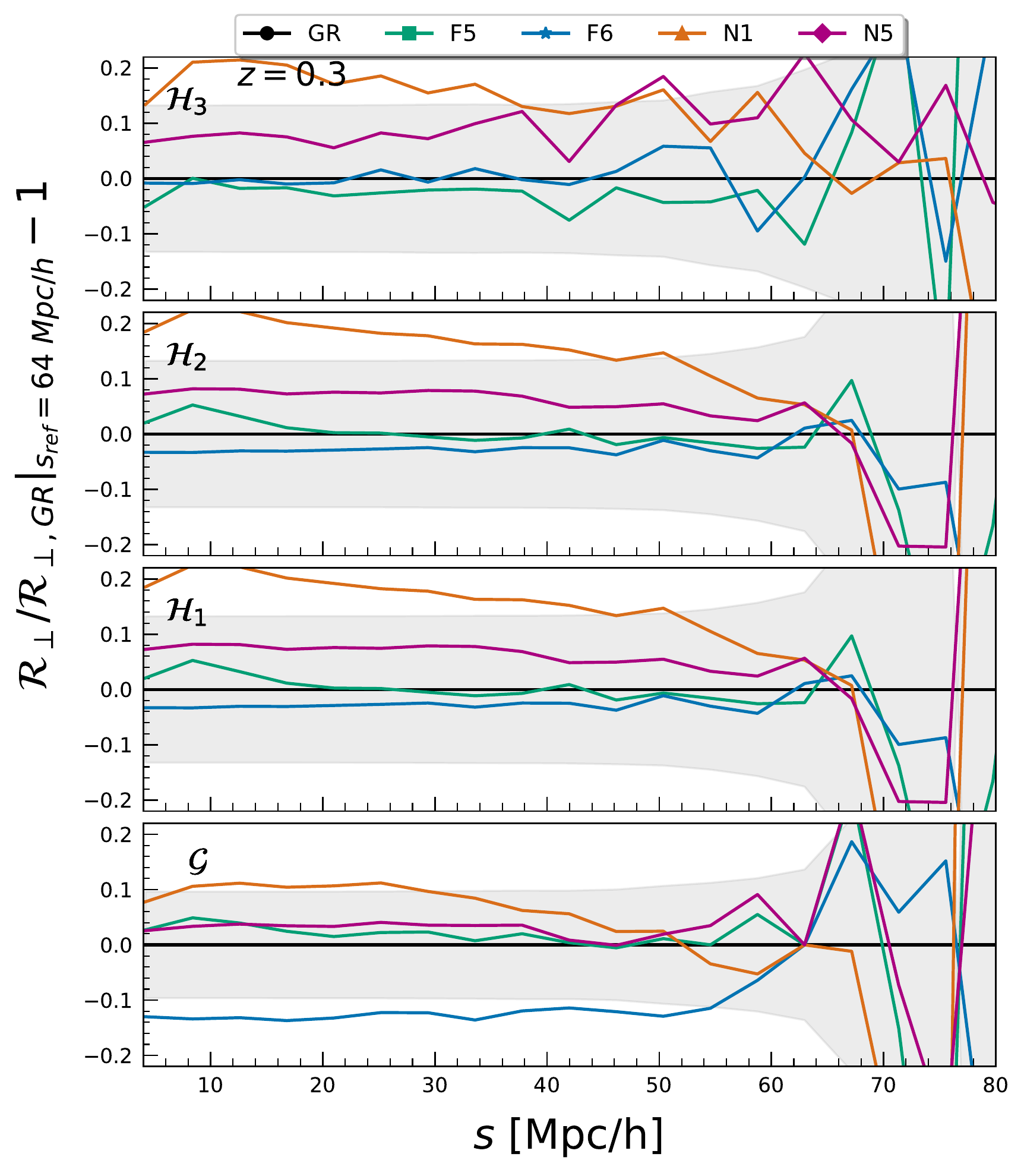}
 \caption{Same as Fig. \ref{fig:HP_ZS_RelClust_xi02} but for the ratios of the redshift-space clustering wedges (parallel and perpendicular) of the 2PCF.}
 \label{fig:HP_ZS_RelClust_xi_w}
\end{figure*}


\section{Relative clustering analysis}
\label{sec:RelClustering}

The above-discussed estimators of the distortion parameter $\beta$, based on the $Q(s)$ and $R(s)$ ratios, depend strongly on the bias, and therefore,
the straightforward interpretation of the clustering is limited by the degeneracy between the growth rate and the bias,
both encoded in $\beta$. Those ratios might also suffer from a theoretical model bias, since they ignore
the effects of the LOS velocity dispersion
\cite[see \eg][]{Bose2017}.
To foster analysis of clustering observables that would be less dependent on the bias model, we
will now consider an estimator inspired by
Arnalte-Mur \etal\ \cite{Arnalte-Mur2017MNRAS}, where a ratio between 2PCFs of different halo
populations at a chosen scale was studied. In that work, the relative clustering ratio, $\calr$, for the monopole of the 2PCF of halo populations was introduced.
Taking the ratio of the two amplitudes of 2PCFs eliminates the linear bias contribution to the first order.
Here we extend that definition to the quadrupole moment and clustering wedges of the 2PCF. This will allow us to look
for signatures of MG encoded also in the higher-order multipole moments of the anisotropy produced by RSD.

The clustering ratio of multipole moments of order $l$, or clustering wedges $w$, $\calr_{l,w}$ is
defined as follows:
\begin{equation}
\label{eqn:clusteringratio}
\calr_{l,w}\left(s, \mathcal{H} \mid \mathcal{H}_{\text {ref }}, s_{\text {ref }}\right)
=\frac{s^{2} \xi_{l,w}(s \mid \mathcal{H})}{s_{\text {ref }}^{2} \xi_{l,w}\left(s_{\text {ref }}
\mid \mathcal{H}_{\text {ref }}\right)}\,\,,
\end{equation}
where $\mathcal{H}_\text{ref}$ is a reference halo (galaxy) sample and $s_\text{ref}$ is a fixed reference (comoving) scale.
The choice of the reference halo/galaxy sample should be made in a way to minimize the effect of noise.
In general, a population with the highest abundance would seem best suited here. However, in the case of data from real galaxy surveys,
an optimal galaxy population could be the one with the best completeness or the largest sky coverage.
Nevertheless, in our calculations we use uniformly selected halo and galaxy samples considered in a distant observer approximation.
Thus, we opt for a simple criterion of the highest sampling rate for our choice of the reference halo population,
i.e. in our case $\mathcal{H}_\text{ref} = \mathcal{H}_1$, of number density $\bar{n}=10^{-3}h^3\text{Mpc}^{-3}$.

Since we deal with only one galaxy sample, it will requite special treatment. The reference and target populations will be identical, so the clustering ratios will reduce to
a normalization of the signal by its amplitude at the reference scale $s = s_\text{ref}$. This will also be the case whenever we compute $\calr$ for the reference halo sample
itself. Therefore, in all
these cases, the relative clustering maximizes the differences between the MG models depending on the reference scale $s_\text{ref}$.
In what follows we explore the behavior of the relative clustering ratios by
considering two different reference scales, i.e. $s_\text{ref} = 16$ and 64 \Mpch. These are the same values as used by \cite{Arnalte-Mur2017MNRAS}. Table \ref{tab:srefclustratios} gives the amplitudes of the multipole
moments (top) and clustering wedges (bottom) of the 2PCF for GR at these reference scales.
The four panels of Fig.~\ref{fig:HP_ZS_RelClust_xi02} show the relative differences in
the clustering ratio for the monopole $\calr_0$ (left-hand panels)
and the quadrupole $\calr_2$ (right-hand panels) for all our halo and galaxy samples taken at the discussed reference scales, upper panels for $s_\text{ref} = 16$
and lower panels for $64$\Mpch.
Similarly, Fig.~\ref{fig:HP_ZS_RelClust_xi_w} displays the relative difference between the clustering ratio
defined for wedges, parallel $\calr_\parallel$ (left-hand panels) and transverse $\calr_\perp$ (right-hand panels), again for $s_\text{ref} = 16$ at the top and $64$\Mpch\ at the bottom.

The analysis of the generalized clustering ratio that we have introduced in Eq.~\eqref{eqn:clusteringratio} leads to interesting conclusions. In general, no single ratio is the `smoking gun' for the MG imprint and various models exhibit beyond-GR signals, if any, in various statistics and for different tracer populations. Looking first at Fig.~\ref{fig:HP_ZS_RelClust_xi02}, for the monopole ratio $\calr_0$ we see that for the \HIII\ halos both F5 and N1 show clear deviations with respect to \lcdm\, of $\sim-5\%$, for both reference scales. Some signal can be also observed for these models in \HII\ for $s_\text{ref}=16$\Mpch, and for N1 only in the \HII\ and \GA\ tracers if $s_\text{ref}=64$\Mpch. Also the F6 model exhibits some departures from GR in this statistic, namely for \HII\ \& \HIII\ for the smallest reference scale, and for \GA\ with the 64\Mpch\ reference. For all other combinations of MG models, tracers and reference scales, the differences in $\calr_0$ with respect to \lcdm\ are mostly within the statistical noise.

The quadrupole clustering ratio, $\calr_2$, seems to be less sensitive than $\calr_0$ as a discriminant of gravity models. For the the 64\Mpch\ reference scale, most of the MG model-population combinations do not show differences from GR that would stand out from statistical scatter, only N1 offers a strong signal at $20<s[$\Mpch$]<30$. The situation is better for $s_\text{ref}=16$\Mpch, where for all the tracer populations the N1 model exhibits significant, of $\sim+10\%$ or larger, deviations from \lcdm\ in this statistic. For \GA, \HI\ and (less significantly) \HII\,, also N5 could be possibly discriminated from GR using $\calr_2$ with the smaller reference scale; in fact, for galaxies the signal is in this case of the same strength for both nDGP models. Some signal is visible also for F6 for the middle tracer populations (\HI\ \& \HII), but its amount of $\sim-2\%$ is hardly outside the statistical errors.

In Figure~\ref{fig:HP_ZS_RelClust_xi_w} we illustrate the differences between MG and GR for the clustering ratios computed for the
parallel and perpendicular redshift-space wedges.
For both wedge types and the two reference scales N1 generally departs from GR by several percent, although there are exceptions for some of the populations where the signal is not significant.
While the mass selection of
\HIII\ seems to be make it possible to distinguish the effects of MG in the F5 model, which departs from GR by up to 5\% up to 60\Mpch, for the denser samples this is no longer possible in general, therefore this behavior is somewhat reverse with respect to N1.
Furthermore, the ratio of the parallel wedge evaluated
for both reference scales
shows that the relative differences of the clustering of N5 increase monotonically with scale, while these appreciable
differences do not appear in the conventional analysis of wedges when comparing with respect to the \lcdm\ model. Next, for the clustering
ratio of the transverse wedge
it is not efficient to use a large reference scale, $s_\text{ref}=64$\Mpch\ to disentangle
the screening effects of modifications of gravity. Although the relative clustering attempts to eliminate the dependence on the linear bias
in the correlations, the F6 model shows a trend that tracks closely the standard model, not being distinguishable even
when the densest halo population is considered. This means there are other features of the structure growth predicted by MG models,
especially in $f(R)$-based models such as F6, that are strongly degenerate with \lcdm\ and that do not allow us to unveil the effects
of modified gravity with RSD, and we cannot even fully remove the issue of the linear bias when using the clustering ratios.

\subsection{Estimation of $\beta$ from clustering ratios}

\begin{figure}
 \includegraphics[width=\linewidth]{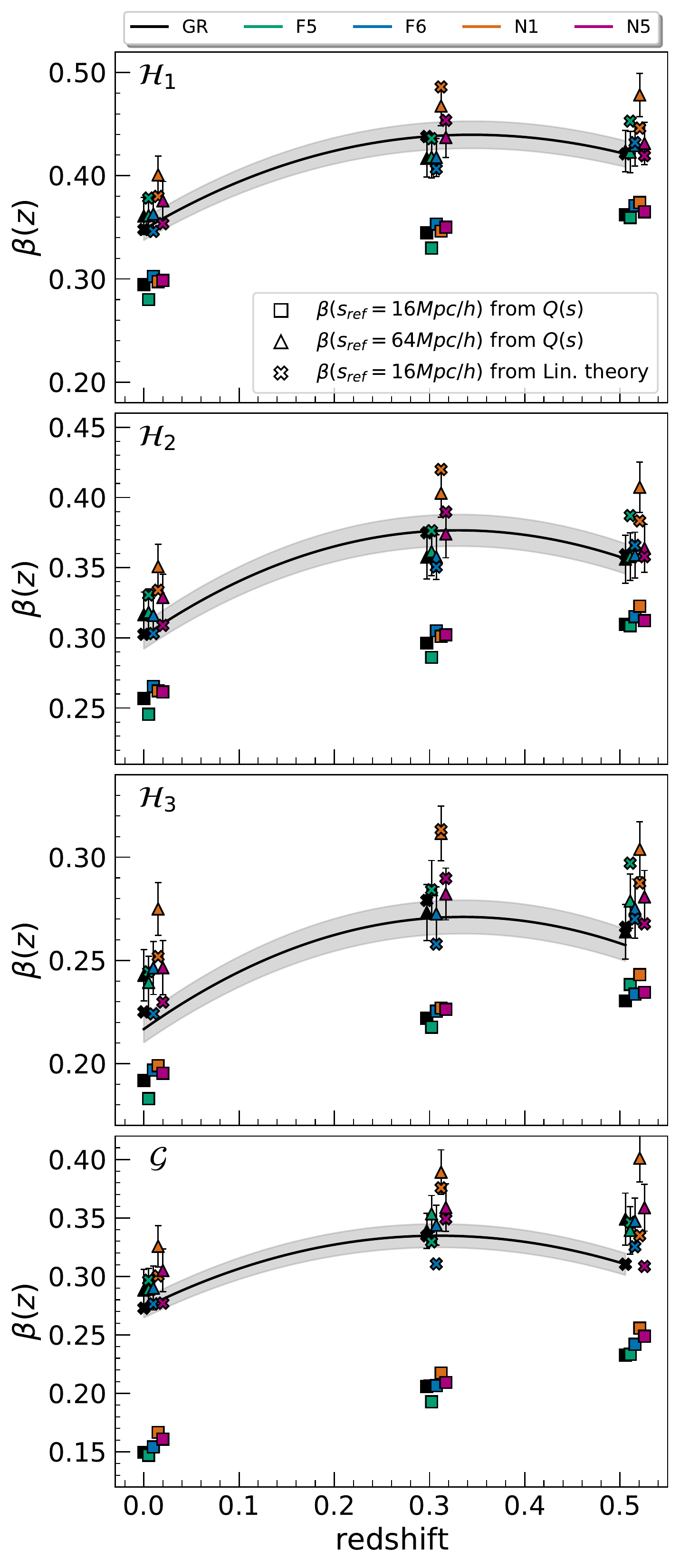}
 \caption{The linear distortion parameter, $\beta(z)$, obtained from the clustering ratios of halos (galaxies). The different
 symbols correspond to different $\beta_\text{ref}$ values used in the $P$ estimator (Eq.~\ref{eq:P_of_s_sref}), as illustrated in the legend. For
 better visualization, the redshift of the MG models has been shifted from
 the mean values $z=0,~0.3$ and $0.5$. The solid line shows the theoretical \lcdm\ prediction computed according to Tinker \etal \cite{Tinker_bias_2010}.
 The error bars represent propagated statistical noise and the gray shaded areas show a $3\%$ error for reference.}
 \label{fig:beta_plot_rel_clust}
\end{figure}

We propose a new estimator for the redshift-space distortion parameter $\beta$ based on
linear theory results and the above-introduced relative clustering ratio,
given
a reference sample and scale. It is defined via the following equation:
\begin{equation}
\begin{aligned}
 P(s)&=\frac{\xi_{DM}(r)}{\xi_{DM}(r_\text{ref})}\frac{\beta_\text{ref}^2(s)}{1+\frac23\beta_\text{ref}+\frac15\beta_\text{ref}^2}\frac{s^2}{s_\text{ref}^2\calr_0(s)},\\
 &=\frac{\xi_{DM}(r)}{\xi_{DM}(r_\text{ref})}\frac{\beta_\text{ref}^2(s)}{R(\beta_\text{ref},s)}\frac{s^2}{s_\text{ref}^2\calr_0(s)}.
\end{aligned}
\label{eq:P_of_s_estimator}
\end{equation}
Here, $R(s)$ is given by Eq.~\eqref{eqn:xi_ratios_multipoles}, and this $P(s)$ estimator is an extension of the $Q$ and $R$ statistics discussed in \S\ref{sec:clustering}.
Although for a theoretical analysis the real space 2PCF can be obtained from Boltzmann solvers that implement HALOFIT \cite{Smith_halofit_2003}, such
as CAMB \cite{Lewis_CAMB_2000ApJ} or CLASS \cite{CLASS_Lesgourgues2011}, we focus instead on the expected values that can be measured from observables in redshift space.
The real-space contribution in Eq.~\eqref{eq:P_of_s_estimator} can be canceled out by evaluating $P(s)\vert_{s_\text{ref}}$, which gives:
\begin{eqnarray}
 P(s)\vert_{s_\text{ref}}&\equiv&\frac{\beta_\text{ref}^2}{R(\beta_\text{ref})}\frac{1}{\calr_0(s_\text{ref})}=\frac{\beta^2}{R(\beta)}
 \label{eq:P_of_s_sref}
\end{eqnarray}
As $P$ can be obtained directly from the measured 2PCF, this provides a natural scaling of the distortion parameter for MG cosmologies, both for
halo and galaxy samples. As the main consequence of implementing the $P$ estimator, we highlight that the accuracy in the resulting
$\beta$ value for a given number density, different from the reference sample, will depend strongly on how well $\beta_\text{ref}\equiv\beta(s_\text{ref},\mathcal{H}_\text{ref})$ is constrained, \ie, the distortion parameter obtained from
the reference sample and scale. Fig.~\ref{fig:beta_plot_rel_clust} shows the linear distortion parameter obtained from
the $P$ estimator, for all MG models and samples, in both reference scales in which the clustering ratio was evaluated,
\ie\, $s_\text{ref}=16$ and $64$\Mpch, as well as for the linear prediction, as indicated by the different symbols. As previously
pointed in \S\ref{sec:RelClustering} the \HI\ population exhibits self-consistent results since it corresponds to the reference sample.

For all tracers, using $\beta_\text{ref}$ for $s_\text{ref}=16$\Mpch\ leads to significantly underestimated $\beta$, when compared to the linear theory.
This effect gets more pronounced at low redshifts, $z\leq0.3$.
The amount of this
bias varies from up to $45\%$ for the galaxy population, down to $15\%$ for \HIII. This underestimation is not surprising, as at 16\Mpch\ the amplitudes of 2PCF are significantly affected by the FoG effect,
and other potential issues connected to small non-linear scales. Since we do not attempt to model the velocity damping term,
our estimation of linear-theory $\beta$ using the $P(s)$ estimator must give a result biased low.
In contrast, using $\beta_{\text{ref}}$ for $s_{\text{ref}}=64$\Mpch\ yields much more self-consistent results.
This is expected, since the impact of the random virialized motions of halos and galaxies onto the clustering
is much weaker at those scales when compared to the weakly non-linear regime at 16\Mpch\ \cite{Okumura2011_RSD_MG,reid2011towards,Scoccimarro_review_2004}.

Assuming that all the relevant effects are modeled precisely, the $P$-based estimator for $\beta$ would foster a significant measurement
of the distortion parameter. In principle, it would offer precision that should allow to distinguish between \lcdm\ and at least the F5 and N1 models.
We can emulate such a situation by using a linear-theory prediction for each model as our $\beta_\text{ref}$.
In that case, we obtain a much better accuracy of the extracted $\beta(z)$ than from the $Q(s)$ measurements. Of course, an actual real-data measurement cannot assume
a value of the parameter that is to be measured. However, our exercise with taking $\beta_{\text{ref}}\equiv\beta_{linear}$
illustrates a theoretical accuracy limit, which one can approach by carefully modeling all the relevant non-linear effects such as velocity damping and scale-dependent galaxy bias.
This will be explored in future work.

\section{Discussion and Conclusions}
\label{sec:discussion_conclusions}

In this paper we have performed a systematic search for modified gravity signals that would be
encoded in various halo and galaxy clustering statistics. Starting from linear theory,
we considered a set of basic predictions, such as halo mass function, linear growth rate parameter,
and effective linear bias. We have employed a set of N-body simulations, the \elephant\ suite, and
used mock halo and galaxy catalogs to construct three halo ($\mathcal{H}_{1,2,3}$) and one galaxy (\GA) samples.
We then proceeded to measure the full 2D two-point correlation function using a distant observer approximation
for redshift-space distortions modeling. Next, we have studied various moments of the full 2D 2PCF and
other related statistics against their potential sensitivity to the underlying gravity model.
Our main focus was on the $z=0.3$ snapshot. This epoch, from all the available snapshots, should be the
closest representation of the BOSS LOWZ ($z\lesssim0.4$) data, the characteristics of which were used to construct galaxy mock catalogs in \elephant.
We can summarize our most important findings as follows.\\\\
{\it Clustering in the real space:}\\
The amplitudes of the matter 2PCF and power spectrum are enhanced in MG models, compared to the GR baseline prediction.
This is a result of a continued action of the fifth-force on the DM fluid. The net effect seen in simulations agrees well with the enhancement
of the linear growth-rate parameter as predicted from the perturbation theory. However, this effect is not translated straightforwardly
to the measured galaxy and halo 2PCFs. We have shown that here the additional factors, like biasing and population selection criteria, affect
the resulting correlation amplitudes. These all add up to produce sometimes counter-intuitive or surprising results.
For example, the N1 model is characterized by an excess over the \lcdm\ amplitude for \HI\ and \HII\ samples, but shows no
significant difference for \HIII\ and \GA. In contrast, the F5 model predicts weaker clustering for the \HII\ sample, while for
all the other populations it is close to GR predictions. In addition, there is no clear trend with MG signal
and the sample density, as for the F5 case our sparsest population \HIII\ can exhibit similarly strong deviations as N1 for \HI.
For all models, the galaxy sample exhibits a clustering amplitude consistent with each other with a typical scatter of 1-2\%.
This result is expected, as by construction the amplitude of the projected correlation function was enforced
when building the HOD mocks. The overall picture emerging from our analysis of the real-space clustering points to significant non-linear behavior
across various samples and models. This emphasizes the need to use carefully designed simulations for predicting
halo and galaxy clustering in MG.\\\\
{\it Halo and galaxy bias:}\\
Our study clearly shows that the linear bias parameter in MG can differ significantly from the GR one for mass-selected samples. The departure from GR of the linear bias at the fixed halo mass has been already indicated by previous studies \cite[see \eg][]{Wyman2013,Achitouv2016PhRvD,Barreira2016,Mirzatuny2013nqa}. Here, we also showed that the effective bias representative of the fixed number-density samples, varies across models
for the same sample selection.
Our results also confirm that for all the studied MG models, as well as for the GR case, the scale dependence of the linear bias
is weak. The resulting relative differences of the bias with respect to \lcdm\ is also relatively flat and only slightly changing with
scale. The F5 model is a clear exception here, however. The deviation of the F5 bias for all but \HIII\ samples
experiences visible non-linear scale dependence. Noticeably, for this $f(R)$ model the bias usually takes values higher than in \lcdm.
In contrast, for the N1 case it is always lower than for GR for all our samples. F6 and N5 show much smaller differences that are typically
not significant, given the scatter. We also studied the redshift evolution of the bias parameter in the range of $0<z<0.5$. We found that the differences from GR are larger at higher redshift, indicating that bias modeling accounting for MG-related physics should be really important for all $z>0$ galaxy data.

In general, we find that the bias of galaxy and halo samples can take significantly different values in MG than in the fiducial \lcdm\ model.
The net effect can have both positive and negative signs and additionally depends on the sample number density.
This indicates that there is an important degeneracy between the enhanced structure formation, as predicted by increased growth-rate of
MG models, and their bias values, different than in GR. This degeneracy needs to be carefully considered in all MG growth-rate measurements employing biased tracers, and therefore, its impact on the growth-rate estimation must be eliminated or at least minimized.\\\\
{\it Clustering in the redshift space:}\\
The iso-correlation contours of the MG 2D 2PCF exhibit visible deviations from the GR case
for the three halo samples. For the case of the galaxy sample the deviations of the contour lines are much milder than in GR. The modulation of MG effects with the changing number density, from \HIII\ to \HI\ samples highlights that
the amplitude of real space clustering varies more in the transverse than the LOS direction.
\begin{itemize}
\item Monopole moment of the 2PCF:\\
We find that the clustering amplitudes of the \HII\ and \GA\ samples are very close to each other when paired consistently for all MG models.
This results in very similar, albeit not identical, trends of departures from GR in those two samples.
For all samples N1 shows a clustering excess in the $\xi_0(s)$ amplitude at all probed scales. The magnitude of this excess is
also correlated with the number density of the sample. The signal for N5 follows qualitatively the same trends, but
with moderately smaller excess amplitude. The $f(R)$ family exhibits quite different behavior. Here typically
the monopole amplitude takes lower values than in the \lcdm\ case. However, the F5 model for the \HI\ sample departs from
this trend
as $\xi_0(s)$ takes here values higher than GR. For nDGP models the observed excess
of the clustering amplitude is relatively flat with scale, while the $f(R)$ family show some weak scale-dependence,
with smaller scales usually showing a stronger signal. We can attribute this latter behavior to the chameleon screening
mechanism that due to its environmental dependence adds additional variability with scale in this class of models.
\item Quadrupole moment of the 2PCF:\\
The analysis of the quadrupole moments yields one clear prediction: only the N1 model shows a signal of deviation from GR that is strong and statistically significant. The net amplitude boost is quite large, and takes up to $20\%$
for all four samples. N5 also shows some deviations, but they are much more minor in the amplitude and hence only marginally significant. Both of our $f(R)$ variants show virtually no signal here, as their quadrupole moments take values that are very close to the fiducial GR case.
\item Clustering wedges:\\
The signal of the MG-enhanced structure formation appears to be better visible in the analysis of the clustering wedges than of the multipoles. The net effect on the amplitude boost of the LOS wedge reaches 10\% and 5\% for N1 and N5 variants, respectively.
The significance of this signal is also prominent, since the statistical uncertainty levels remain comparable
with the one for the monopoles. This shows promise of using the LOS clustering wedges for rendering
stronger constraints on the nDGP class of models. On the other hand, for the perpendicular sky direction,
the galaxy sample was much less sensitive to MG effects. The effect was still prominently present
in the halo samples, but now contained only to small non-linear scales.
Thus, we argue that the perpendicular clustering wedge is not well suited for testing growth-rate deviations.
Interestingly, for both parallel and perpendicular wedges the amplitudes for $f(R)$ samples
are close to the \lcdm\ case, deviating by no more than $\sim2\%$. These indicates that the wide-angle
integration over the 2D 2PCF results in a statistic that is rather insensitive to the $f(R)$-model effects.
\item Linear distortion parameter $\beta$:\\
Our measurements of $\beta(z)$ performed with the quadrupole-based estimator $Q(s)$ clearly illustrate that the beyond-GR modified bias and the modified growth-rate are combined in a way that is not straightforward to model nor account for it at best.
In the case of N1, where the increased $f(z)$
is accompanied by a decrease of the galaxy/halo bias, the net effect
on $\beta(z)$ is always positive. This is reflected in the estimated $\beta$ values that clearly for all redshifts and samples lie systematically above the \lcdm\ case. The N5 model should, in principle, exhibit a similar combined trend, but the net effect is small given our statistical uncertainties. Therefore, our results here are only offering a hint in this direction.

For both $f(R)$ variants we find, however, that the resulting distortion parameter for all samples and redshifts
is always very close to the fiducial GR case. The weak and non-significant result for F6 is actually expected, as
this model in general is characterized by only very weak departure from the GR structure formation.
For the F5 model, the increased bias
conspires against the enhanced growth rate. In this case the total effect brings $\beta$ values much closer to the \lcdm\ scenario. This might be somewhat surprising given the fact that the F5 variant in the linear regime deviates rather significantly from GR.

The overall picture emerging is: while the quadrupole growth rate estimator offers good sensitivity to the nDGP models, it performs very poorly as a method to differentiate between GR and F5 or F6.
\end{itemize}
{\it Measurements of the clustering ratios:}\\
The relative clustering ratios promise to be less susceptible to the growth rate-bias degeneracy, as revealed by our analysis of the estimators based on correlation function moments. This is related to the fact that in the ratio, the effect of the first-order linear bias contribution is reduced.
\begin{itemize}
\item {Ratios of multipole moments:}\\
The analysis revealed that the smaller of our two reference scales, \ie\ 16\Mpch, in general fosters bigger differences of the MG clustering ratios versus the \lcdm~ case, than for $s_\text{ref}=64$\Mpch.
For halos, on top of the notoriously strong signal of N1,
only the monopole ratio taken from the \HIII{} population contained a significant signal for another model, F5 in this case.
We find however a very promising result for the galaxy population and the quadrupole ratios. In this case, both N1 and F5
are characterized by a clear and significant signal attaining nearly a $\sim 10\%$ difference from GR with $\sim 2\sigma$
significance. The results taken for the reference scale at 64\Mpch\ are characterized by a much bigger scatter. This has diminished the significance of nearly all the signals, with two notable exceptions for the monopole ratios of \HIII\ and \GA\ samples.
The \HIII\ population of N1 and F5 models presents a $\sim 5\%$ deviation, but with a marginal $1\sigma$ significance. Among the exciting results for the galaxy sample, we find a clear signal of $5\%$ for the F6 model with statistical
significance $\geq 1.5\sigma$ which emerges at $s\simlt 40$\Mpch\ for $\calr_0$.
The \GA\ sample yields also marginally significant deviation for N1 at $3\%$ level.
\item {Ratios of clustering wedges:}\\
The landscape is largely similar for the case of the clustering wedge  ratios. Although the MG effects are now
typically smaller at $s_{\text{ref}}=16$\Mpch\ than at 64\Mpch, but the significance behaves actually inversely.
From the combination of the scales and models, we showed that the  ratios of the clustering wedges can actually accommodate
significant deviations for N1 and F5 MG variants. Looking at smaller scales, also the F6 and N5 models manifest departures from GR. Here, especially exciting looks a large $\sim12\%$ deviation pertaining for scales up to 50\Mpch\ that the F6 model
shows in the galaxy sample. This signal maintains a nearly $1\sigma$ significance for all the scales.
\item {Estimation of $\beta(z)$:\\
Using the definition of the clustering ratios, we have formulated a new estimator for $\beta$ based on the ratio of two monopoles in redshift space, \ie\ $\calr_0$.
Applying this estimator for ratios taken at $s_{\text{ref}}=64$\Mpch\ results in a similar performance as in the case
of the standard $Q$-based $\beta$-estimator. For the small reference scale, $s_{\text{ref}}=16$\Mpch, the distortion parameter values are largely underestimated, which indicates that the non-linear effects at those scales, both in the velocity and the density field,
are significant. Ignoring this leads to a severe bias in the $\beta$-parameter estimation. Using additional information,
in our case the value of the linear-theory predicted $\beta_{\text{ref}}$,  greatly improves the accuracy and the performance of the new
estimator. Now, for most redshifts and samples the results for all MG  variants are clearly separated; this illustrates the power of more thorough modeling. Using the linear theory prediction allows one to break the bias--growth-rate degeneracy. This exercise yields therefore a theoretical maximal sensitivity of the $\calr_0$-based estimator, which could be achieved in the case of accurate small-scale modeling. This modeling would potentially include scale-dependent non-linear bias and velocity damping.}
\end{itemize}

The enhanced structure formation fostered to a various degree by all MG scenarios we considered leads
to clear predictions in the linear regime. However, the presence of highly-nonlinear fifth-force screening mechanism,
\ie\ the Vainshtein and the chameleon effects, in general increase significantly the total degree of non-linearity in such scenarios.
Our analysis of the redshift-space clustering of four different samples across five variants of structure formation scenarios
 clearly confirms that there is rich potential in using such clustering statistics both for testing the self-consistency of GR,
as well as for searching for alternative MG signals. We have unveiled MG signals present in the various statistics,
such as multipole moments, clustering wedges and clustering ratios across our halo and galaxy samples. Similar results for some
of the statistics considered here were previously found by other authors
\cite[see \eg][]{Jennings_RSD_MG_2012,Marulli_2012B,Wyman2013,Arnalte-Mur2017MNRAS,Hellwing2017,Aguayo2019oxg,Liu2021weo_fRDGP,DESI_ELEPHANT}.

The deviations from GR-baseline of the various clustering statistics should manifest themselves as  measurable differences in the linear growth-rate, $f$. A common procedure to measure it involves extracting the $f\sigma_8$ product from the best-fit model of the data. This combination is treated as a convenient single parameter to be compared across different data sets and different models. What is actually measured, however, is $\beta\sigma_8^{G}$, a product of the linear distortion parameter normalized by a given galaxy sample linear variance, $\sigma_8^G$, taken at $s=8$\Mpch. Therefore, marginalization over $b\sigma_8$ needs to be done,
in order to obtain an $f\sigma_8$ measurement.
Our analysis is indicating potential dangers of this procedure if the resulting constraints on $f\sigma_8$ are to be used for testing gravity. Due to highly-nonlinear
behavior of tracer bias in Modified Gravity, especially in the case of chameleon $f(R)$ theories, such naive marginalization may lead
to a biased $f\sigma_8$ result. Only when the scale dependence of MG, GR bias and other sources of degeneracy are well mapped, a comparison of $f\sigma_8$ for different gravity models and galaxy samples can be regarded as self-consistent and free of severe systematics.

The effective galaxy/halo  bias in the MG models we studied here can
differ from the GR-baseline by at least $\sigma_b/b_{MG}\simeq \pm0.05$. Such variability, if unaccounted for, will add
a systematic effect into the growth-rate measurement. A robust study based on high-resolution simulations for each inquired MG model is needed to chart the growth-rate-bias degeneracy and add such modeling
into $f\sigma_8$ determination. In case such  robust modeling is not yet available for a given MG model, we advocate a safer approach consisting of using the linear distortion parameter instead. In addition, $\beta(z)$ can be readily measured for each galaxy sample separately and used for constraining and comparing GR and MG consistently
within the given data-set. The trade-off is that the resulting $\beta$ parameter is less sensitive to modified growth-rate in the models
where the galaxy bias can be modified in a non-linear way, such as $f(R)$. When dealing with models that accommodate more
predictive bias modifications, such as the nDGP class, $\beta(z)$ already could be used for obtaining robust constraints on
strongly deviating variants, such as N1 tested here. If the nuisances are controlled and viably modeled,
this parameter can yield competitive constraints on MG models from the current and especially future spectroscopic galaxy surveys.

Our study also indicates that smaller scales contain much more constraining power when using RSD for testing gravity. While this is a relatively well-known and
appreciated fact for the case of the standard \lcdm{} analysis, it is even more vital to tap the small-scale potential for conducting
competitive and stringent MG tests with the use of clustering data. A natural next step  to exploit this potential would be
a systematic study of the non-linear small-scale bias and the effects of galaxy pairwise velocity dispersion in the context
of RSD and MG physics. Such a program is already underway, and we will present its results in a forthcoming study.

\section*{Acknowledgments}

This work is supported via the research project ``VErTIGO'' funded by the National Science Center, Poland, under agreement no 2018/30/E/ST9/00698. The authors are grateful for support from the Polish Ministry of Science and Higher Education (MNiSW) through grant DIR/WK/2018/12. WAH \& MB also acknowledge support of the National Science Center, Poland under a grant no. UMO-2018/31/G/ST9/03388.


\bibliographystyle{h-physrev-fix}
\bibliography{ELEPHANT_RSD_clustering-new.bib}

\label{lastpage}
\end{document}